\documentstyle[prb,aps,psfig]{revtex}
\begin{document}
\preprint{Bicocca/FT-01-26  October 2001}
\draft

\title
 {Critical universality and  hyperscaling revisited for Ising models\\
  of general spin  using extended high-temperature  series \\}

\author{P. Butera\cite{pb} and M. Comi\cite{mc}}
\address
{Istituto Nazionale di Fisica Nucleare\\
Dipartimento di Fisica, Universit\`a di Milano-Bicocca\\
 3 Piazza della Scienza, 20126 Milano, Italy}
\date{15 October 2001}
\maketitle
\begin{abstract}

We have extended through  $\beta^{23}$ the high-temperature expansion
of the second field derivative of the susceptibility for  Ising models
of general spin, with nearest-neighbor interactions, on the simple cubic 
and the body-centered cubic lattices. Moreover the expansions for the 
nearest-neighbor correlation function, the  susceptibility and  the second 
correlation moment  have  been extended up to $\beta^{25}$. 
Taking advantage of these new data,  we can improve the accuracy  of  direct 
estimates of critical exponents and of  hyper-universal combinations of 
critical amplitudes such as  the renormalized four-point coupling $g_r$ or  
the quantity usually denoted by $R^+_{\xi}$. In particular, we obtain 
$\gamma=1.2371(1)$, $\nu=0.6299(2)$, $\gamma_4=4.3647(20)$, $g_r=1.404(3)$ 
and  $R^+_{\xi}=0.2668(5)$.    We have used a variety of series extrapolation 
procedures and, in some of the analyses, we have assumed that the leading 
correction-to-scaling exponent $\theta$ is universal and roughly known.  
We have also verified, to  high precision, the validity of the  hyperscaling 
relation and of the  universality property both with regard to the  lattice 
structure and to the value of the spin.
\end{abstract}

 \widetext

\section{Introduction}
The numerical study of the critical properties of the spin-$S$ 
 Ising models with  nearest-neighbor interactions 
 had an important historical role in the chain of arguments leading
 to the modern formulation of the universality 
hypothesis\cite{univ,dg,wilson,pok,dc,mefrmp} 
  for the critical 
phenomena and, in particular,  to 
the concept of universality class. 
It was in a  study of 
the susceptibility $\chi(\beta;S)$ for the general-spin Ising models 
on the face-centered-cubic (fcc) lattice 
 by high-temperature (HT) expansions
through $\beta^6$,  
  that C. Domb and M.F. Sykes\cite{dosy}  first pointed 
out that the exponent $\gamma$, which characterizes the divergence of $\chi$, 
 was roughly  independent 
 of $S$ and guessed for it a universal  
``daltonian'' value $\gamma=5/4$ .
Later on, when longer series\cite{camp,camp2,moore} both for 
the fcc and for other lattices 
 were derived, a weak 
dependence of the exponents $\gamma$ and $\nu$ 
on  $S$ emerged from   the HT analyses, but
 was soon correctly ascribed  
to the occurrence of non-analytic ``confluent corrections 
to scaling'' (CCS) rather than to a failure of the universality. 
In those years the existence of  CCS  had been
  inferred by various authors both from  the numerical analysis of  HT 
series\cite{fishbur,wort,saul} and   from phenomenological
fits\cite{kouv,grey,balz} to high precision experimental data for 
 some  systems close to criticality.
 Eventually the status of
 the CCS was 
 more firmly established\cite{wegner} in the context of
 the Renormalization Group (RG) theory\cite{wilson}.
It was therefore recognized  very early  that the accurate 
 determination
of the critical exponents  in numerical or experimental studies
 and, as a consequence, the feasibility of 
stringent verifications both of the universality hypothesis
 and of the scaling and hyperscaling relations require a close
control over the CCS.
For many years, however,     in two and in three 
dimensions,   HT series were  available only for  few observables and,
 generally, they were barely sufficient to conjecture the presence 
 of CCS, but  definitely too short 
 to make a numerically accurate  
discussion of these models possible\cite{camp,moore,gabriel,chakra}.
 Still presently, the expansions  of
  $\chi(\beta;S)$ and of the second moment of the correlation 
function $\mu_2(\beta;S) $  on the simple-cubic (sc) 
lattice, for spin $S > 1/2$, can be found  explicitly 
 in the literature\cite{camp2} 
only up to the order $\beta^{12}$.  
The   data files 
 by R. Roskies and P.M. Sackett\cite{sack}  made 
an extension of these series through $\beta^{15}$ feasible
 for the sc and the bcc lattices,  
 but did not drastically change the situation. 
 On the fcc lattice, the HT series initially derived through
 $\beta^{12}$ in  Ref.\cite{camp2}, 
 were later extended  in Ref.\cite{mckest} 
 to order $\beta^{14}$.
 Fortunately, in the case of the 
body-centered-cubic (bcc) lattice, 
 decisive progress  occurred  already two decades ago,
 with the computation by B.G.Nickel\cite{ni21} of 
 expansions for  $\chi(\beta;S)$ and  $\mu_2(\beta;S)$ through $\beta^{21}$. 
(To our knowledge, only the series for $S=1/2,1,2,\infty$ 
were published\cite{nr90}.) 

By allowing to some extent for the leading CCS, 
 the first modern analyses of the extended bcc  
series\cite{ni21,nr90,z81,fhr,fhrb,geor,georth} 
improved significantly
 the accuracy in the verification of universality 
with respect to the magnitude of the spin.
Moreover,  
 in the mentioned studies 
(as well as in later analyses\cite{oths} mainly devoted 
 to the $S=1/2$ case), 
the central estimates of the susceptibility 
and the correlation-length exponents were  
reduced up to  $\approx 1\%$ with respect to 
 the values $\gamma=1.250(3)$ and $\nu=0.638(2)$, 
  initially guessed  in Ref.\cite{dosy} and later confirmed\cite{fisre,moo} 
by various studies. 
This development also contributed to settle\cite{dc}
   a long-standing controversy raised by the results of 
Refs.\cite{fishbur,moo,bahy,bin,bfre}, which stimulated the studies of Refs.
\cite{ni21,nr90,z81,fhr,fhrb,geor,georth,oths,ga,carg,rohy,nish,rehr,mck,g86}, 
on the validity of hyperscaling and,  more generally, on the consistency 
of the results from the HT analyses  with the corresponding RG
 estimates \cite{dc,rgeps,rgfd,murnick,zinn,itzy,cardy,kleb},
  either in the $\epsilon$-expansion 
approach\cite{rgeps} or in the fixed-dimension perturbative 
scheme\cite{rgfd,murnick}. 

One should also note that  for the second field-derivative of the 
susceptibility $\chi_4(\beta;S)$ 
and for the nearest-neighbor correlation function $G(\beta;S)$,  
 the published data  are even less abundant. 
 On  the sc lattice, series for $\chi_4(\beta;S)$ 
 can be  derived  from the   data files
  of Ref.\cite{lw}
  up to order $\beta^{14}$, and up to $\beta^{10}$   
 from the data of Ref.\cite{bin} for the bcc lattice.
 On the fcc lattice, 
 series  for $\chi_4(\beta;S)$  are available\cite{mck}  through $\beta^{13}$.
  For general spin, only expansions\cite{camp2,mckest} of 
 $G(\beta;S)$   through  $\beta^{14}$ on the fcc lattice 
have been published.
 A summary of the HT expansions available until now  for the  
 Ising models of  general spin appears in Table \ref{tab1}.

 We have been pursuing  
 a long-term  project 
 to improve the algorithms and the codes for  HT expansions
 in  two-dimensional\cite{bc2d} 
and  in three-dimensional\cite{bcsd,bcgn,bc3d,bcon,bc23} lattice spin models, 
 keeping up with the 
steady increase of computer performances, and  periodically
 updating the numerical analyses whenever we could 
 significantly extend the series. 
By using an appropriately  renormalized
 linked-cluster method\cite{bin,lw,lcm},
 we have now added from four up to thirteen terms to  
the HT expansions for various observables of the general spin-$S$ 
Ising models on  the sc and the bcc lattices. In this paper we shall examine
 the expansions of $\chi(\beta;S)$ and $\mu_2(\beta;S)$  up 
to order $\beta^{25}$ 
 and of  $\chi_4(\beta;S)$ up to  $\beta^{23}$ on both lattices.
These data have been derived  by slightly improving  
 the  thoroughly tested 
code which recently produced \cite{bc23} our series  
through $\beta^{23}$ 
for $\chi(\beta;1/2)$ and $\mu_2(\beta;1/2)$ on 
 both lattices.  
The extension of the series is by far the hardest part of this work,
 but we will not enter enter here into the details of our procedure.
 To give an idea of the required computational effort, it will
suffice to mention that  our improved 
codes take minutes of CPU time on a COMPAQ
Alpha XP1000 (500 MHz) single-processor workstation to
reproduce the known series through  $\beta^{21}$, whereas several days 
are necessary to add the following four orders. 
From the graph-theoretical point of view, it is 
the expansion of $\chi_4$ through $\beta^{23}$ which involves 
 the most laborious part
 of the calculation: in the simplest vertex renormalized 
expansion  scheme\cite{lw} it would require 
the generation and the evaluation of 
 over $10^9$ topologically
 inequivalent graphs.
 However, devising a 
 careful strategy of in-depth
 renormalizations, the expected size of the calculation has  been reduced  
 by at least two orders of magnitude.
 On the other hand, from a purely computational standpoint, 
 the calculation of the sc lattice constants for
  the second moment of the correlation function is 
the most  demanding part of the job in terms of CPU-time.

 The correctness of our codes is 
 ensured  by numerous internal consistency checks, as well as by their
 ability to reproduce established results already available 
 in simpler particular cases,
 such as the square-lattice  two-dimensional spin 1/2 Ising model or 
the one-dimensional spin-$S$ Ising models. 
 Of course,   our codes also  reproduce 
the old computation  of Ref.\cite{ni21} 
for $S=1,2,\infty$   on the square and the bcc lattices,  
and, as far as there is overlap, also the recent computation
 of Ref.\cite{cam} for $S=1/2$ on the bcc lattice.

 Using this vast 
library of partially new high-order series data 
 and in particular  our significantly 
extended series for $\chi_4(\beta;S)$,   
 we can   resume from a vantage point  
 the very accurate studies performed on series $O(\beta^{21})$ for 
 $\chi$ and $ \mu_2$ in  
 Refs.\cite{ni21,nr90,z81,fhr,fhrb}
  and  present an even  more  extensive and 
detailed  survey of the 
critical behavior for the spin-$S$ Ising models. 
In spite of the  remarkable advances achieved 
 by the calculations of Ref.\cite{ni21,nr90,z81,fhr,fhrb,geor}  
 which  removed away from the 
foreground the universality and the hyperscaling issues, 
 further extensions of the 
 HT data still remain of great interest. 
 They are instrumental in the continuing efforts 
to gain  a higher accuracy in the estimates 
of the  critical parameters and, more generally, to 
perform   more stringent 
 tests of hyperscaling and of universality, 
with respect both  to  the value of the spin 
and to the lattice  structure.
 These are certainly welcome results, since    
 it is fair to say that the actual verification of such basic properties  
 is still only moderately accurate, although no doubts persist 
anymore about their validity.  
  Of course, one must be aware that the computational complexity of 
the calculation of higher-order series coefficients
 grows much faster than the  
  precision in the evaluation of the critical parameters that 
can be obtained  from them by the presently available numerical tools. 
  Therefore the higher-order
 computations should 
  be  accompanied also by an 
 effort to improve the techniques of analysis or, at least, by a careful 
 comparison of the results obtained by a variety of methods. 

The paper is organized as follows.
In the next Section, we set our notations and definitions. In Sec.III 
we state the assumptions underlying our analysis and its aims. 
 The numerical procedures we have used, namely the 
modified-ratio methods  introduced in Ref.\cite{z81} or the 
differential approximant methods\cite{guttse,gutasy}, as well as 
 the corresponding results of the series analysis,  
are discussed in Sec. IV-VIII.
In Sec. IX we compare our estimates with those of the most recent
 literature.
 The last few Sections  present our results for  the 
 critical amplitudes of the observables that have been expanded 
 and for some (hyper)-universal combinations\cite{aha} of these amplitudes. 
 In order to make our analysis completely reproducible and to provide   
a convenient source of data
 for further work, without overburdening this paper, 
 we have collected into a separate report\cite{bcunp}, available on request, 
  the complete expansions  of the nearest-neighbor correlation function, of
 the susceptibility,  
  of the second moment of the correlation function
  and of the second field 
derivative of the susceptibility for spin
$S=1/2,1,3/2,2,5/2,3,7/2,4,5,\infty$,   
 on the sq, sc and the bcc lattices.

\section{ The spin-$S$ Ising models}

  The spin-$S$ Ising models are defined by  the Hamiltonian:

\begin{equation}
H \{ s \} = -{\frac {J} {2}} \sum_{\langle \vec x,{\vec x}' \rangle } 
s({\vec x})  s({\vec x}') 
\label{hamilt} \end{equation}

where $J$ is the exchange coupling, and
  $s(\vec x)=s^z(\vec x)/S$  with  $s^z(\vec x)$  a  
classical spin  variable at the
lattice site $\vec x$, taking the $2S+1$ values 
$-S, -S+1, \ldots,S-1, S$.    
 The sum runs over  all nearest-neighbor pairs of  sites. 
We shall consider expansions in the usual HT variable $\beta=J/k_BT$ called
``inverse temperature'' for brevity. 

 In the high-temperature phase, the basic observables are the 
connected $2n$-spin correlation functions. Here we shall 
limit our  study to quantities related to the  
 two-spin correlation functions $\langle s(\vec x)  s(\vec y) \rangle_c$
 and to the four-spin correlation functions 
$\langle  s(x)  s(y) s(z)  s(t)\rangle_{c}$.

In particular, we shall consider 
the nearest-neighbor correlation function   

\begin{equation}
G(\beta;S)= \langle s(\vec 0)  s(\vec \delta) \rangle_c=
 \sum_{r=0}^\infty h_r(S) \beta^r
\end{equation}
  where $\vec \delta$ is a nearest-neighbor lattice vector.

The internal energy per spin 
is defined in terms of 
$G(\beta;S)$ by 
\begin{equation}
U(\beta;S)=- \frac{qJ} {2} G(\beta;S)
\end{equation}
where $q$ is the lattice coordination number.

The specific heat is  the temperature-derivative of the internal 
energy at fixed zero external field

\begin{equation}
C_H(\beta;S)/k_B = \frac{q\beta^2} {2} \frac {dG(\beta;S)} {d\beta} 
\end{equation}

In terms of $\chi(\beta;S)$, the zero-field reduced susceptibility,

\begin{equation}
\chi(\beta;S) = \sum_{\vec x} \langle s(0)  s(\vec x) \rangle_c = 
\sum_{r=0}^\infty c_r(S) \beta^r 
\label{chi} \end{equation}

and  of  $ \mu_{2}(\beta;S)$, the second  moment 
of the correlation function,

\begin{equation}
 \mu_{2}(\beta;S)=\sum_{\vec x} \vec x^2 \langle s(0)  s(\vec x) 
\rangle_c = 
 \sum_{r=1}^\infty d_r(S) \beta^r, 
\end{equation}

 the  ``second-moment correlation length'' $\xi(\beta;S)$ is defined  by 

\begin{equation}
 \xi^{2}(\beta;S)= \frac  {\mu_{2}(\beta;S)} {6\chi(\beta;S) }=
\sum_{r=1}^\infty t_r(S) \beta^r.
\end{equation}

The second field-derivative of the
susceptibility $\chi_{4}(\beta;S)$ is defined by

\begin{equation}
 \chi_{4}(\beta;S)=  \sum_{x,y,z}
\langle  s(0)  s(x) s(y)  s(z)\rangle_{c}=
 \sum_{r=0}^\infty e_r(S) \beta^r. 
\label{chi4}\end{equation}

 Notice that these  
definitions ensure  the existence of a non-trivial limit as 
$S \rightarrow \infty$.

\section{Assumptions and aims of  the series analysis }
 In the  universality 
 class of the spin-$S$ Ising models,  the asymptotic 
behavior of the susceptibility
 as $\beta \rightarrow \beta^{\#}_c(S)$ from below, 
 is expected to be

\begin{equation}
 \chi^{\#}(\beta;S)
\simeq C^{\#}(S)\tau^{\#}(S)^{-\gamma}\Big(1+ a^{\#}_{\chi}(S)
\tau^{\#}(S)^{\theta} + \ldots
+ b^{\#}_{\chi}(S)\tau^{\#}(S) + \ldots \Big)
\label{conf}\end{equation} 

where $\tau^{\#}(S)= 1- \beta/\beta^{\#}_c(S)$ is the reduced 
inverse temperature. We have introduced here the 
superscript ${\#}$ which stands for either sc or bcc, as appropriate,
and will be used hereafter only when useful.
Eq.(\ref{conf}), often called  the Wegner expansion\cite{wegner},
  specifies how 
the dominant scaling behavior, characterized by the 
universal critical exponent $\gamma$ 
 and by the   critical amplitude 
$C^{\#}(S)$,
  is modified by analytic and nonanalytic confluent corrections 
to scaling(CCS) in a close 
vicinity of the critical point.  
 The leading nonanalytic CCS 
 is characterized by a universal exponent  $\theta$ and by an  
  amplitude $a^{\#}_{\chi}(S)$.
The  critical amplitudes 
$C^{\#}(S), a^{\#}_{\chi}(S),
 b^{\#}_{\chi}(S)$,  as well as  the
 inverse critical temperature $\beta^{\#}_c(S)$, are 
nonuniversal:   namely they  depend on the spin $S$ and 
 on the lattice structure,
as stressed  by the notation. 
The analogous asymptotic behaviors of the 
correlation length 

\begin{equation}
 \xi^{\#}(\beta;S)
\simeq f^{\#}(S)\tau^{\#}(S)^{-\nu}\Big(1
+ a^{\#}_{\xi}(S)
\tau^{\#}(S)^{\theta} + \ldots
+ b^{\#}_{\xi}(S)\tau^{\#}(S) +\ldots \Big)
\label{confxi}\end{equation} 

of the specific heat

\begin{equation}
 C_H^{\#}(\beta;S)/k_B
\simeq  A^{\#}(S)\tau^{\#}(S)^{-\alpha}
\Big(1+ a_C^{\#}(S)
\tau^{\#}(S)^{\theta} +\ldots
+ b_C^{\#}(S)\tau^{\#}(S) + \ldots \Big)
\label{confc}\end{equation}

and of  $\chi_{4}(\beta;S)$

\begin{equation}
 \chi^{\#}_{4}(\beta;S)
\simeq -C^{\#}_{4}(S)\tau^{\#}(S)^{-\gamma_4}\Big(1+ a^{\#}_{4}(S)
\tau^{\#}(S)^{\theta} +\ldots
+ b^{\#}_{4}(S)\tau^{\#}(S) + \ldots \Big)
\label{conf4}\end{equation}  
as well as of the other singular observables, 
 are characterized  by 
 different critical
 exponents and by different (nonuniversal) critical amplitudes 
$f^{\#}(S), a^{\#}_{\xi}(S),$ etc., but all contain 
the same leading confluent exponent $\theta$.
  Notice that we have freely chosen in   
equations (\ref{conf})-(\ref{conf4})    between the conventions  of
 Ref.\cite{liufi} and those of Ref.\cite{aha}, 
since the notation for the amplitudes is not yet  completely standardized.

 Usually the exponent $\gamma_4$ is expressed in terms
of $\gamma$ and of $\Delta$, the ``gap''
 exponent  associated with the critical 
behavior of the higher field-derivatives of the free energy,
as follows: $\gamma_4= \gamma+2\Delta$. Here $\Delta=\beta+\gamma$ and 
$\beta$ denotes the magnetization exponent only in this formula and 
in the scaling relation to be quoted before eq.(\ref{hyp2ma}).
 From RG calculations\cite{rgeps,rgfd,murnick,zinn,itzy,cardy,kleb,guida,zeq} 
it is  expected 
 that  $\theta \simeq 0.5$ for the Ising universality class.

For later use, we observe that, if the  singularity
  closest to the origin in the complex $\beta$ plane is the critical 
singularity, then  eq.(\ref{conf})
 implies  the following asymptotic behavior for the expansion 
coefficients $c^{\#}_n(S) $ of $\chi^{\#}(\beta;S)$

\begin{equation}
c^{\#}_n(S)= 
C^{\#}_{\chi}(S)\frac {n^{\gamma-1}} {\Gamma(\gamma)} \beta^{-n}_c(S)
\Big [1+ \frac { \Gamma(\gamma)}
{\Gamma(\gamma-\theta)} \frac{a^{\#}_{\chi}(S)} {n^{\theta}}+
O(1/n)\Big ]
\label{asycoe}\end{equation}
In the Ising universality class, 
the expected $O(1/n^{2\theta})$ contributions in eq.(\ref{asycoe}) 
 should be practically 
degenerate with the analytic $ 1/n$ corrections. 
For bipartite lattices, the higher-order corrections in eq.(\ref{asycoe})
 include 
terms $O(1/n^{1+\gamma-\alpha})$ with 
alternating signs, which
 reflect the presence\cite{fam,syfi,wowy} 
of a weak ``antiferromagnetic'' singularity 
at $\beta=-\beta_c^{\#}(S)$ with exponent $1-\alpha$.
   Analogous formulae for the 
asymptotic behavior with respect to the 
order  can be written  for the expansion coefficients  
 $t^{\#}_n(S)$ of  $\xi^{\#}(\beta;S)^2$ and 
 $e^{\#}_n(S)$ of $\chi^{\#}_4(\beta;S)$.

In terms of $\chi$, $\xi$  
and $ \chi_{4}$, a ``hyper-universal'' combination of critical 
 amplitudes denoted by $ g_r$ and usually called   the
``dimensionless renormalized
 coupling  constant'',   can be defined in $d$ dimensions 
by  the limiting  value  of the ratio

\begin{equation}
 g^{\#}(\beta;S)\equiv - \frac{3 v^{\#}  \chi^{\#}_{4}(S;\beta)}
{16 \pi\xi^{\#}(S;\beta)^d \chi^{\#}(S;\beta)^2}
\label{gr}
\end{equation}

as $ \tau^{\#}(S) \rightarrow 0+$.
 Here $v^{\#}$ denotes the volume per lattice 
site (in 3 dimensions $v^{sc}=1$ 
 and $v^{bcc}=4/3\sqrt3$) and   the normalization factor 
 $\frac{3} {16\pi}$ is chosen in order 
to match the conventional field theoretic definition 
of $g_r$\cite{zinn}.  We shall call $g^{\#}(\beta;S)$ 
the ``effective renormalized coupling constant''
at the inverse temperature $\beta$.

By eqs.(\ref{conf})-(\ref{conf4}),  
$g^{\#}(S;\beta)$  behaves  as 

\begin{equation}
  g^{\#}(\beta;S)
\simeq  g_r\tau^{\#}(S)^{\gamma+d\nu-2\Delta}\Big(1+ a^{\#}_{g}(S)
\tau^{\#}(S)^{\theta}+ \ldots \Big)
\label{confg}\end{equation} 
 when $ \tau^{\#}(S) \rightarrow 0+$, with

\begin{equation}
  g_r
=  - \frac{3 v^{\#}  C^{\#}_{4}(S)}{16 \pi f^{\#}(S)^d C^{\#}(S)^2}
\label{ampg}\end{equation} 

The  Gunton-Buckingham\cite{gunt,fishbuck,rig,glja} inequality 
\begin{equation}
 \gamma + d\nu  -2\Delta \geq 0 
\label{gunt}\end{equation}

 together with the Lebowitz\cite{lebo} inequality 
$\chi^{\#}_4(\beta;S) \leq 0$, 
 ensures that $ g^{\#}(\beta;S)$ remains bounded and non-negative 
 as $\tau^{\#}(S) \rightarrow 0+$. 
 The  vanishing of $ g^{\#}(\beta^{\#}_c-0+;S)$   is a sufficient condition
for Gaussian behavior at criticality, namely 
for the vanishing of the four-spin and of the higher-order 
connected correlation functions. In lattice field theory
language, this corresponds 
to the ``triviality'' \cite{wilson,fernandez}, numerically 
 observed when $d= 4$ and 
proved when $d> 4$, for the continuum field 
theory defined by the   lattice model (with ferromagnetic 
couplings\cite{afc,gri}) in the critical
 limit. If   the  inequality (\ref{gunt}) holds as an equality 

\begin{equation}
 \gamma + d\nu  -2\Delta = 0
\label{hyp}\end{equation}

(called the `hyperscaling relation'), if there are no logarithmic
 corrections to the scaling behavior\cite{bctn,bct}  and if
 $\chi^{\#}_4(\beta;S)$ is nonvanishing,  we have:

\begin{equation}
  g^{\#}(\beta;S)
\simeq  g_r\Big(1+ a^{\#}_{g}(S)
\tau^{\#}(S)^{\theta}+ \ldots
\Big)
\label{confgg}\end{equation} 

namely the `` effective coupling'' 
$ g^{\#}(\beta;S)$  tends to a universal nonzero 
limiting value  $g_r$ as $\tau^{\#}(S) \rightarrow  0^+$.

Using the Essam-Fisher\cite{essa,rush,rushl} 
 scaling relation $\alpha+2\beta+\gamma=2$, 
 eq.(\ref{hyp}) can be rewritten as a 
relation\cite{jose,josew} between $\alpha$ and $\nu$ 
  
\begin{equation} 
\alpha= 2-d\nu .
\label{hyp2ma}\end{equation}

As a consequence, 
 also  the  following combination of the critical 
amplitudes $A^{\#}(S)$ and $f^{\#}(S)$

\begin{equation}
R^+_{\xi}=(\frac{\alpha A^{\#}(S)} {v^{\#}})^{1/d}f^{\#}(S)
\label{confrx}\end{equation}

 is hyper-universal, as  pointed out\cite{stauff} by 
Stauffer, Ferer and Wortis. 

The  ratios
$a_{\xi}/a_{\chi}$ and $a_{4}/a_{\chi}$, $a_{C}/a_{\chi}$ etc.,  
of the amplitudes of the  leading CCS
 are less studied\cite{aha}, but not less interesting,  
 universal \cite{ferer,chought,baber} critical observables.

In the rest of this paper we
 shall employ our HT series to estimate the critical parameters
defined by eqs.(\ref{conf})-(\ref{conf4}), 
 (\ref{confgg}) and (\ref{confrx}),  to check the 
validity  of eqs.(\ref{hyp}) 
 and (\ref{hyp2ma}) and  of the universality property with respect to the 
 value of the spin and to the lattice structure.

In the actual numerical analysis of finite-order HT expansions, 
the presence of the CCS will generally become
 manifest\cite{camp,ni21,nr90,z81}
 by small apparent violations both of the hyperscaling 
relations and of the universality properties, namely by  
a weak apparent  dependence of the universal quantities  on the 
lattice structure and on the value $S$ of the spin. We shall point out this
 fact  by explicitly indicating, in our 
 numerical estimates of the universal quantities, the value of the 
 spin and the lattice structure  
for the  series used in the analysis. For instance, $\gamma^{bcc}(S)$ will 
 denote the numerical estimate of the universal 
exponent $\gamma$  obtained from the series 
 $\chi^{bcc}(\beta;S)$.
 This notation will help  to emphasize  how small the mentioned 
 effects of apparent nonuniversality 
 are reduced  if the  HT expansions  can be pushed to  a   sufficiently 
high order, provided that  the numerical tools of the analysis
 can, at least approximately,  allow  for the leading CCS. 

 Part of our  analysis 
will rely upon the main assumption 
 that  the exponent $\theta$ of the leading CCS is  universal 
 and  roughly known. 
 A recent  accurate  
RG recalculation of universal critical data\cite{guida,zeq} 
 predicts the value $ \theta=0.504(8)$ in the 
fixed-dimension perturbative approach, 
 while  within the $\epsilon$-expansion scheme, the updated 
 estimate 
is $ \theta=0.512(13)$.  
In the rest of this paper,  we shall adopt 
as a reference value the fixed-dimension RG  estimate
$ \theta^{ref}=0.504(8)$, when  computing  the central values of 
  critical parameters   by procedures biased with  
 $\theta$.
 Even if one has   no compelling reason to suppose that the uncertainty of
the RG prediction of $\theta$ is largely underestimated,  
 (but this possibility is advocated in Ref.\cite{case}),
  the reliability of the $\theta$-biased 
    analyses presented here will be greater  
  whenever  their results are not too sensitive 
to the precise value of  $\theta$. 
In the following Sections, it will be clear   that,
   in most cases, 
 we can tolerate an uncertainty of this exponent  
 even several times  larger than  above indicated. 
  We will show that most of our  
 estimates biased with $ \theta^{ref}=0.504(8)$ 
will be  compatible also with  
  higher values,  
 such as  $\theta=0.52(3)$, proposed in Refs.\cite{nr90,geor}, 
 $\theta=0.53(2)$   in Ref.\cite{has} or even $\theta=0.54(3)$
 from Refs.\cite{fhr,fhrb,zinnfish,bail,blh}.
We can add that, both our direct HT evaluation of $\theta$ and 
the part of our series analysis
 which is not biased with the
value of the exponent $\theta$,  
will be completely consistent with the above assumption.

If appropriate, we shall provide 
detailed  information on the $\theta$-biased numerical
 results reported in our tables for a given quantity $P$,  by  indicating 
  together with the central estimate, also
 the derivative $\partial P/ \partial \theta$ 
 evaluated  at the reference value chosen for 
$\theta$. Similarly, in the cases where the parameter estimates are biased 
 with the value 
of a critical inverse temperature $\beta_c^{\#ref}(S)$ 
 and/or of a critical exponent, for instance $\gamma$, 
 we shall report   the corresponding derivatives 
  $\partial P/ \partial \beta_c$ and/or $\partial P/ \partial \gamma$
 computed at the specified reference values.
As an example, for the critical amplitude of the 
 susceptibility $C^{\#}(S)$,  our final estimate can   be read as
\begin{equation} 
 C^{\#}(S)(error) 
+(\partial C^{\#}(S)/\partial \beta_c) (\beta_c -\beta_c^{ref}) 
+(\partial C^{\#}(S)/\partial \gamma) (\gamma- \gamma^{ref}).
\label{biasedeq}\end{equation}
 Here both the estimate and its derivatives 
are evaluated for sharp values of 
$\beta_c= \beta_c^{ref}$ and of $\gamma= \gamma^{ref}$ 
 and the {\it error} attached to the first term
 does not allow for the  uncertainty 
of the bias parameters.  Since the above
  expression describes how the central estimate of 
$ C_{\chi}^{\#}(S)$ changes under  small variations 
of the bias parameters,  
 comparisons with previous results in the literature, often 
based on slightly different  assumptions, are  made  straightforward.

 As a final general remark, it is worth to mention
  that, due to the higher coordination number of the lattice,
  the bcc series approach their asymptotic 
 structure eq.(\ref{asycoe}) generally faster than the sc series. 
For this reason, the bcc series 
  are usually observed to yield  more accurate estimates 
of the critical parameters than the corresponding sc series with the same 
number of coefficients  and are 
often   said to have a greater  ``effective length''. 
 This fact will be confirmed here, and will be one of the reasons to 
 draw our final best estimates from the analysis of the bcc series. 
 Nevertheless, the sc lattice series 
 remain very interesting, in particular because 
the non-universal informations obtained from them 
 are  directly  comparable to the  data from simulation studies,
traditionally performed  on the sc lattice. 
It is also interesting to notice that,  for both lattices,
 the asymptotic behaviors set in  more 
slowly in the most widely studied $S=1/2$ case. This is not surprising 
since the number of degrees freedom per site is proportional 
to the magnitude of the spin.
 A slower convergence
  is observed also for the higher moments of the correlation 
 function, since in their construction
 larger weights are given to the correlations between
 farther sites for which the expansions are effectively 
 shorter. As a consequence, on both the sc and the bcc lattices, 
  the  expansions of  $\xi$ shows a  slower convergence than those of
 $\chi$.

 Basing on the assumptions above indicated, our analysis will   
 aim to exhibit, within the family of the spin-$S$ Ising models,
 some consequences of the universality property, of  
the scaling  and of the hyperscaling laws
  for the correlation functions, for the exponents and for 
various universal  combinations of critical amplitudes. 
In particular our accurate verification of the universality property
  will strengthen the justification 
of  a technique 
advocated long ago by J. Zinn-Justin\cite{z81} 
and  independently by 
 J.H.Chen, M.E.Fisher and B.G.Nickel\cite{fhr} 
to improve the precision in the computation 
of the universal critical parameters of the Ising model.
 These authors  argued that numerical study 
  should address 
appropriate families of spin models parametrized by a continuous
 auxiliary variable and   belonging to  the same 
universality class as the Ising model.
For specific values of this variable it is possible to
  select representative models 
 for which the amplitudes of the leading confluent corrections  to scaling 
are negligible and, 
as a consequence,  the determination of the universal critical 
parameters can be more accurate.
 By relying on a similar   prescription,  we will also  obtain very 
accurate estimates of some universal critical parameters.

\section{ Estimates of the critical points}

In this Section  
 we shall  examine the HT expansion of 
the susceptibility in zero field, for several values of the spin $S$ 
on the sc and the bcc lattices. The 
series coefficients of the susceptibility generally show
 a very  smooth dependence on the order of expansion
 and  a relatively fast  approach to their asymptotic forms.
 Therefore they are   best suited to an 
 accurate determination of the critical temperatures.
 The estimates so obtained  will also be adopted 
 to bias the calculation of the critical exponents and  amplitudes.

As we have already argued in our previous study\cite{bc23} 
of the HT series for the $S=1/2$ case,
 the modified-ratio  method introduced by
J. Zinn-Justin\cite{z81}, (see also\cite{guttse}), 
 can lead to 
 estimates of the critical inverse temperatures with an accuracy comparable
 or  sometimes 
 higher than the traditional differential approximant (DA) methods.
Perhaps, the potential of this tool has not been properly appreciated, 
because so far
it could not be used with series  long enough.

The method consists in evaluating  
$\beta^{\#}_c(S)$ from the approximant sequence
\begin{equation}                       
(\beta^{\#}_c(S))_n = (\frac {c_{n-2}c_{n-3}} {c_{n}c_{n-1}})^{1/4}
 exp[\frac { s_n+  
s_{n-2}} {2 s_n( s_n- s_{n-2})}]  
\label{betazinn}\end{equation}

with 

\begin{equation}
 s_n=(ln(\frac{c_{n-2}^2} {c_{n}c_{n-4}})^{-1}
+ln(\frac{c_{n-3}^2} {c_{n-1}c_{n-5}})^{-1})/2 .
\label{essen}\end{equation}

 Since the expected value of $\theta$ is very nearly $1/2$,  
by using eq.(\ref{asycoe}),
the asymptotic behavior of the approximant sequence can be  
 expressed as follows:

\begin{equation}
(\beta^{\#}_c(S))_n =\beta_c^{\#}(S)(1 - 
\frac {\Gamma(\gamma)}
{2\Gamma(\gamma-\theta)}
\frac{\theta^2(1-\theta) a^{\#}_{\chi}(S)} {n^{1+\theta}}
 +O(\frac{1} {n^{2}}))  
\label{asbetazinn}
\end{equation}

It is interesting to observe also that, 
 if $\theta=1/2$,
  the coefficient of the 
 $O(1/n^2)$ correction in eq.(\ref{asbetazinn}) is equal 
to $(a^{\#}_{\chi}(S))^2$ times a 
  very small positive factor. Moreover,
  the coefficient of the $1/n$-term in 
 eq.(\ref{asycoe}) enters into eq.(\ref{asbetazinn}) 
only at next higher orders. 
Since $a^{\#}_{\chi}(S)$  is  expected to be  
 small (though generally not negligible),  these remarks help
 to understand how this method works 
 and why it  is much more efficient than 
 the conventional ratio prescriptions.
We can say that eq.(\ref{asbetazinn}) provides an estimate
 of the leading ``finite-order effects '', namely of the 
corrections due to using series of finite length $n$.
 These are strictly analogous to the well known ''finite-size effects''
 which have to be carefully considered to improve the data 
 from simulations of finite systems. At the orders
of expansion presently available eq.(\ref{asbetazinn}) has already 
a reasonably accurate quantitative meaning. 
  
   Although  devised specifically 
to deal with  the expected structure of the singularities, 
 the procedure we have sketched 
is  {\it unbiased}: namely  
 no additional accurate information on other 
critical parameters must be used 
together with the series in
 order to get the estimate sequence.
 However, at the present orders of expansion,  
 the $n-$dependence of  $(\beta_c(S))_n$
is not saturated
 and, for  sufficiently large $n$,  
the successive  estimates   
  show an evident residual   trend, very nearly linear on a 
 $1/n^{1+\theta}$ plot, as expected from eq.(\ref{asbetazinn}). 
 Small  odd-even oscillations are superimposed to 
the main monotonic trend as a 
consequence of the above mentioned antiferromagnetic singularity  
(see the comments to eq.(\ref{asycoe})).
These observations suggest that  one can  do something 
better than taking the highest-order available 
term  of the sequence eq.(\ref{betazinn}) 
 as the final estimate  of $\beta^{\#}_c(S)$. 
 The most obvious improvement consists in using
 the assumed known value of $\theta$  to fit the asymptotic
 behavior of the sequence and in taking the 
extrapolated value of the sequence as a better estimate of $\beta^{\#}_c(S)$.
 As usual,  one should  separately extrapolate  to large 
$n$ the odd and the even subsequences  
 of $(\beta_c(S))_n$, in order to deal properly also 
with the oscillations due to the antiferromagnetic singularity. 
Our extrapolation will be based 
on the successive pairs of terms 
 in the approximant subsequences.
 Eventually  a further minor adjustment of the results 
might be performed by  a second (purely visual) extrapolation 
 in order to allow also for a very small 
residual curvature of the plots due to the higher 
corrections in eq.(\ref{asbetazinn}). 
For instance, in the $S=1/2$ case on the sc lattice, the 
highest-order estimate 
from the  extrapolation of the last  pair 
 in the odd-approximant subsequence is $0.22165646$. 
In order to allow for the small residual curvature 
of the extrapolation sequence, this figure should probably be 
 slightly reduced, to yield our
  final (and very conservative) estimate $\beta^{sc}_c(1/2) = 0.221655(2)$.

 The set of our  estimates for $\beta^{\#}_c(S)$ is reported 
in Table \ref{tab2}.
The errors we have indicated are  small multiples (2-4) of the 
differences between the 
extrapolations of the two highest-order pairs of terms in the odd
 subsequences. In the same Table, we have also reported 
$\partial \beta^{\#}_c(S)/\partial \theta$ evaluated 
at  $\theta=\theta^{ref}$. 
 As shown by our data, the above  mentioned uncertainty 
in the value of $\theta^{ref}$  
turns out to be unimportant in the whole procedure,   
 because it  contributes  only a small fraction  
of the final uncertainty of the estimates.

\begin{figure}[t]
\centerline{\hbox{
\psfig{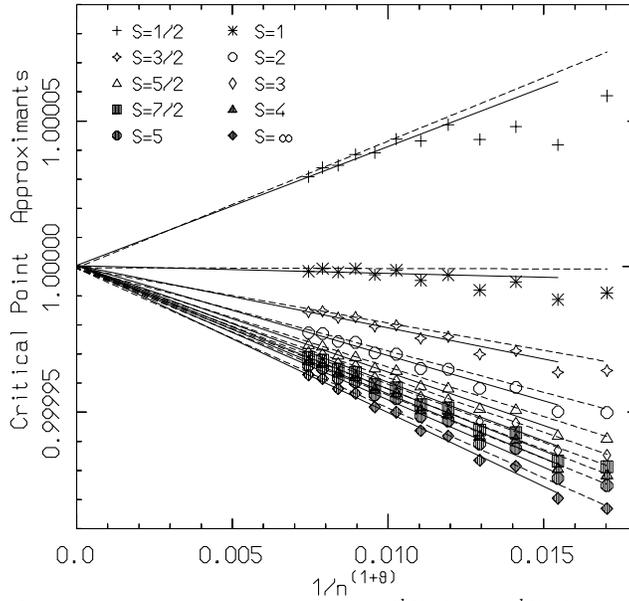}}}
\caption{The ``normalized''  modified-ratio  approximant 
 sequences $(\beta^{bcc}(S))_n/N^{bcc}(S)$  
 for the critical inverse temperature of the spin $S$
 Ising models on the bcc lattice, plotted  
versus $1/n^{1+\theta}$, with $\theta=\theta^{ref}=0.504$.
 They are obtained from eq.(\ref{betazinn}) 
using the coefficients $c_n^{bcc}(S)$ 
of the susceptibility series for the bcc lattice. 
In order to fit into a single figure the 
sequences for different values of the spin, 
  each sequence  has been normalized by the 
average $N^{bcc}(S)$ of the critical 
inverse temperatures obtained extrapolating separately the even 
and the odd subsequences. We have indicated by continuous lines the 
extrapolants of the odd subsequences, based on the last odd pair of
 approximants, while the dashed lines indicate the extrapolants of the  
 even subsequences, based on the last even pair of approximants.}
\label{fig1}
\end{figure}

In  order to give an idea of the qualitative features of the method,  
 for each value $S$ of the spin examined in this study, 
we have plotted  in Fig.\ \ref{fig1} the corresponding ``normalized''
 approximant sequence
 $(\beta_c^{bcc}(S))_n/N(S)$ vs. $1/n^{1+\theta}$.  
 We have taken the average
 of the extrapolated values of 
the even and the odd subsequences as  the normalization factor $N(S)$, 
 introduced only to make the various plots easily  
 comparable and  conveniently fit all of them into a single figure. 
We have drawn as continuous lines the extrapolants of the 
last odd pair of terms in the sequences,
 whereas  the dashed lines indicate 
 the extrapolants
 of the last even pair. The  difference between the extrapolated values 
of the odd and the even normalized subsequences, which is generally
  very small (for instance it is $\approx  10^{-6}$ in the bcc 
lattice case and at most 
four  times as large in the sc case), provides a first 
rough indication that the oscillating 
 corrections due to the antiferromagnetic singularity
give only a small contribution to the uncertainty of the results. 
 The final relative errors reported in Table \ref{tab2} are generally 
much larger.

If we refer to  eq.(\ref{asbetazinn}), the plots in Fig.\ \ref{fig1}  
strongly suggest that 
$a_{\chi}^{bcc}(1/2)<0$, whereas 
 $a_{\chi}^{bcc}(S) > 0$ for $S \geq 2$.    
Since  $|a_{\chi}^{bcc}(1)|$  and $|a_{\chi}^{bcc}(3/2)|$ are 
 very small,  we cannot yet be completely sure about their sign.
The smallness of these confluent corrections 
is confirmed  observing that  
$\vert \partial \beta^{bcc}_c(1)/\partial\theta \vert$
 and $\vert \partial \beta^{bcc}_c(3/2)/\partial\theta \vert$ 
are much smaller than for the other values of $S$.

 Simple model series with a  structure specified by eq.(\ref{asycoe})
 (including the antiferromagnetic oscillating 
corrections) can mimic  rather accurately 
 the behavior of the 
spin-$S$ Ising series for sufficiently high orders. Therefore numerical 
 experimentation with these model series  can  give us 
some intuition on the virtues and the limitations 
 of the modified-ratio method and help to assess its accuracy. 
 These tests add further confidence  on our estimates
 of the  relative error of $\beta_c$.
On the other hand, it may take series significantly longer
 than those presently available to determine 
$a_{\chi}(S)$   with a precision better than 
 a few percents, since the slopes
 of the approximant sequences provide only ``effective'' values of these
amplitudes due to the residual influence of the higher-order corrections.
 Actually, the relative uncertainty of  $a_{\chi}(S)$ can be  larger, 
particularly so if its absolute value is very small.

\begin{figure}[b]
\centerline{\hbox{
\psfig{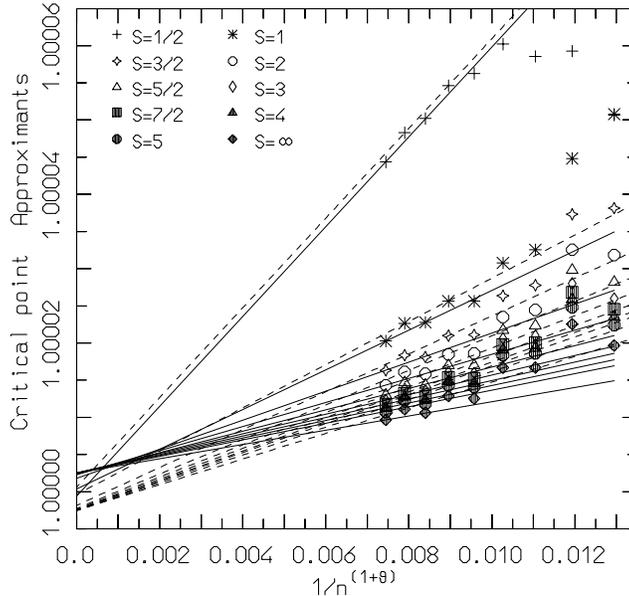}}}
\caption{
The same as in Fig.\ \ref{fig1}, but for the ``normalized''  
modified-ratio  approximant sequences 
$(\beta^{sc}(S))_n/N^{sc}(S)$ formed from 
the  coefficients $c_n^{sc}(S)$ of the 
susceptiblity series for the sc lattice.}
\label{fig2}
\end{figure}

 The sc lattice series have been studied in the same fashion and the 
 results are illustrated by Fig.\ \ref{fig2}.
The main difference with respect to the bcc case is that  all
approximant sequences
 are decreasing, so  that $a_{\chi}^{sc}(S)<0$ for all $S$. 
It is also clear that, for this lattice, the 
 rate of convergence of the approximant sequences to their asymptotic 
behavior is distinctly slower than in the bcc case.

 In order to gain further confidence in the  estimates by the 
modified-ratio method,
 we must  confirm at least their main features  
also by numerical tests of a 
different nature or involving different assumptions, thus
 reducing the probability of being misled by only apparent convergence.
We have therefore  performed  
 also   a more
traditional {\it unbiased} analysis by first 
and second-order inhomogeneous DA's  yielding values 
of the critical inverse temperatures  
 in essential agreement, to within their uncertainties, 
with those obtained from 
the modified ratio-method. In the case of the bcc lattice, 
  for spin $S=1/2$, the highest-order available DA estimates
 are slightly larger than the estimates from the modified ratio-method. 
Nevertheless, the estimates  from DA's using $r$ 
series coefficients show a  slowly 
decreasing trend as $r$  increases.  
 For  $S>2$ the highest-order  DA 
estimates are slightly smaller than the corresponding  results 
from the modified-ratio method, 
 but the estimate sequences show 
an increasing trend.
If we make the reasonable assumption\cite{liufi} that also for DA's 
 the dominant finite-order corrections  are proportional to the  amplitudes
 of the leading nonanalytic corrections to scaling,
these features of the results can be simply explained 
by the  pattern of signs and sizes of these amplitudes  
previously observed in  the  analysis.
 Taking account of these trends and performing some purely visual 
 extrapolation of the DA estimate sequences, 
 we can  reconcile the DA and   the 
  analyses by the modified-ratio method . 
 We shall not report in Table \ref{tab2} the DA results, but  simply 
 quote here  the average of the highest-order DA 
data for a few values of $S$.    
 The presence  of residual 
trends in  the sequence of DA estimates will be indicated by    
asymmetric uncertainties roughly 
 corresponding to the range of possible extrapolations. 
 For instance, from second-order unbiased DA's, we obtain 
 $\beta^{bcc}(1/2)=0.157376^{(+1)}_{(-4)}$, $\beta^{bcc}(1)=0.224655(2)$, 
$\beta^{bcc}(3/2)=0.265640(2)$,   
$\beta^{bcc}(2)=0.293255(2)$ 
and $\beta^{bcc}(\infty)=0.435082^{(+4)}_{(-1)}$. 

 For   the same reasons as above indicated, in the sc lattice case,
  the values of $\beta^{sc}_c(S)$ 
obtained from the DA's are generally slightly 
larger than those suggested by the modified-ratio  analysis. 
This upward shift is more pronounced in the case $S=1/2$, in which we obtain
$\beta^{sc}(1/2)=0.221665^{(+2)}_{(-10)}$, whereas
 for higher values of $S$ the differences from the results by 
modified-ratio method 
are much smaller, for example:  
 $\beta^{sc}(1)=0.312870(3)$, $\beta^{sc}(3/2)=0.368660(3)$
and $\beta^{sc}(\infty)=0.601271(3)$.

 In conclusion,  our modified-ratio method (biased with $\theta$) 
 and the unbiased DA estimates  of the critical 
inverse temperatures  on the sc and bcc lattices are consistent and  
compare fairly well  with, but sometimes are more accurate than
 those already available  in the literature 
for  a few values of the spin and also reported   in the table.
Wider discussion of other  estimates by different methods in 
 the $S=1/2$ case\cite{oths,adler,bcon,blh,bst,land,stau,salm,ito,tama},
 can be found in our previous paper\cite{bc23} and in  recent 
reviews\cite{dpl,blu,labi} of MonteCarlo  simulations
 and other studies of spin models.
It is  interesting to mention at this point that the two most extensive 
  simulations on the sc lattice, 
by a static\cite{bst} and by a kinetic\cite{ito} method,  
yield the estimates $\beta^{sc}_c(1/2)=0.2216546(10)$ and  
$\beta^{sc}_c(1/2)=0.2216595(15)$, respectively, which agree 
 only within two standard deviations.

\section{Modified-ratio estimates of the critical exponents} 

The  modified-ratio methods can  lead also
 to fairly good estimates\cite{bc23} 
 of the exponents $\gamma$ and $\nu$. 
 Let us first focus on the calculation of the exponent $\gamma$  to 
recall  the prescription of Refs.\cite{z81,guttse}.   
 An analogous procedure can be used for  other exponents.
 For each value of $S$, we form 
 the approximant sequence
 \begin{equation}
 (\gamma(S))_n = 1 + \frac{2( s_n+  s_{n-2})} {( s_n- s_{n-2})^2}
\label{gammazinn}\end{equation}
where  $s_n$ is still defined by eq.(\ref{essen}) 
in terms of the expansion coefficients $c_n(S)$  of $\chi(\beta;S)$.

  Using eq.(\ref{asycoe}),  
we can compute the 
asymptotic behavior of the sequence $(\gamma(S))_n$ 
as follows

\begin{equation}
(\gamma(S))_n=
\gamma(S) -  \frac {\Gamma(\gamma)}
{\Gamma(\gamma-\theta)}\frac{\theta(1-\theta^2)a^{\#}_{\chi}(S)} 
{n^{\theta}}+ O(\frac{1} {n})
\label{gammazinnas}\end{equation}

 If $\theta=1/2$, the  $1/n$ term in eq.(\ref{gammazinnas})
 has a coefficient equal to $(a^{\#}_{\chi}(S))^2$ times a small
positive factor.
 The higher-order corrections contain powers of  $\gamma$. 
As a consequence, for the Ising model, the first important correction is 
$O(1/n^{1+\theta})$ and,  in general, the convergence of the 
sequence eq.(\ref{gammazinn})  will be slower 
whenever the exponent under study is  $\gg 1$. 
This the case of $\gamma_4$ and,  
actually, we have observed that, at the presently available orders,
 this procedure is not convenient  for estimating $\gamma_4$,  
whereas a directly biased variant, 
to be described in the next Section, is more successful.  
On the other hand, for the calculation
 of the specific-heat exponent $\alpha$, 
 there are  difficulties of a different nature:  
  the critical singularity is very weak and  the 
 number of nonzero coefficients of the HT expansion of $C^{\#}_H(\beta;S)$ 
is still too small. Because of that,  we  have not been  
 able to  improve by  modified-ratio methods the accuracy of 
the current {\it direct } HT estimates\cite{bcsd,gusc,bha} of $\alpha$.

For sufficiently large $n$, 
the   sequence of approximants defined by eq.(\ref{gammazinn})   
 is very nearly linear on a 
 $1/n^{\theta}$ plot.
 Therefore, arguing like in the previous section, 
 we are led to  improve our estimates by 
extrapolating  the odd (or even) subsequences 
linearly in  $1/n^{\theta}$.  
The higher-order 
corrections for the exponents   are expected to be  more important 
 than in the calculation of $\beta_c$  
 and this  reflects into a larger uncertainty  of the 
extrapolation  procedure.
Just like in the formula for $\beta_c$,
 the limiting value of the approximant sequence 
  is asymptotically 
approached from above, if  the amplitude $a^{\#}_{\chi}(S)$ 
of the leading nonanalytic confluent correction to scaling(CCS) is negative, 
 or from below,  if it is positive. 

\begin{figure}[t]
\centerline{\hbox{
\psfig{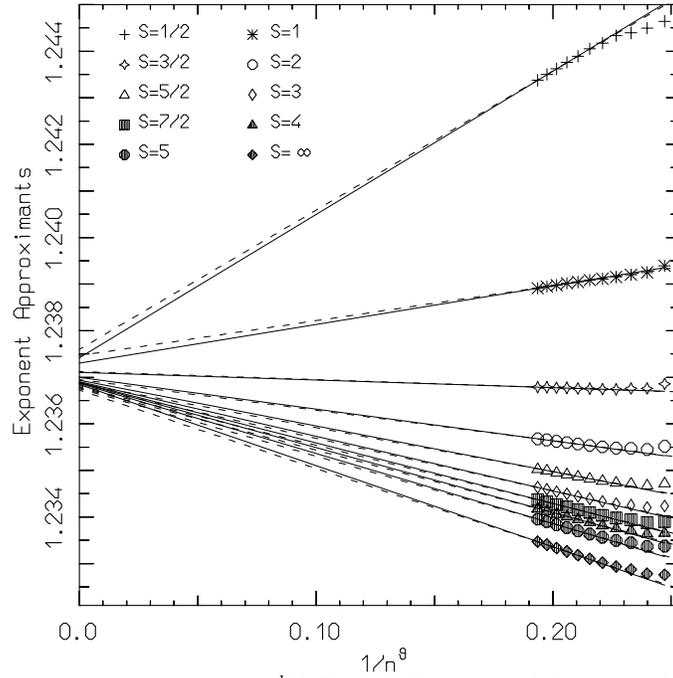}}}
\caption{The modified-ratio  approximant sequences $(\gamma^{bcc}(S))_n$ 
of the susceptibility critical exponents
 for various values of the spin $S$,  
plotted vs. $1/n^{\theta}$, with $\theta=\theta^{ref}=0.504$.
  They are obtained using  eq.(\ref{gammazinn}) from the susceptibility 
series coefficients $c^{bcc}_n(S)$. 
 For each value of the spin, we have  
indicated by a continuous line the 
extrapolation to large $n$ of the sequence, 
linearly in $1/n^{\theta}$, based 
on the last odd pair of terms
$\{(\gamma^{bcc}(S))_{23}, (\gamma^{bcc}(S))_{25}\}$.   
 A dashed line indicates the extrapolation 
based  on the previous
 odd pair $\{(\gamma^{bcc}(S))_{21}, (\gamma^{bcc}(S))_{23}\}$
}
\label{fig3}
\end{figure}

\begin{figure}[b]
\centerline{\hbox{
\psfig{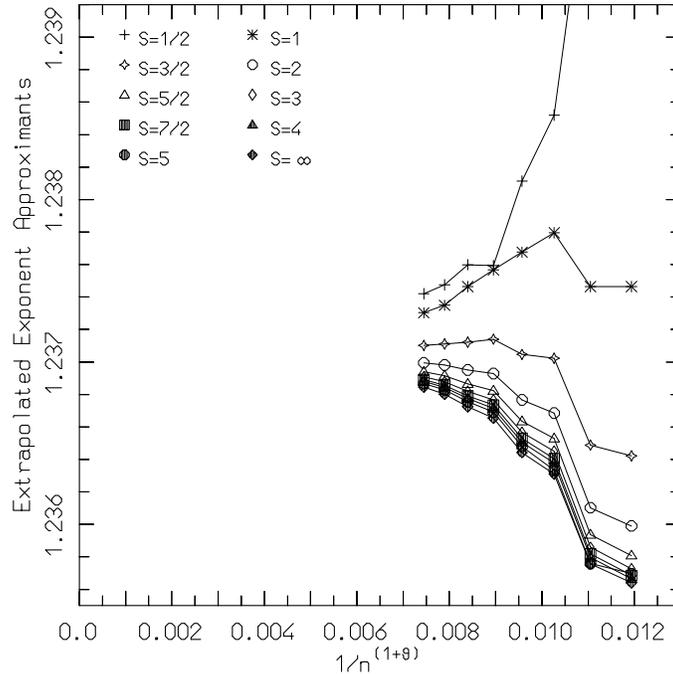}}}
\caption{The sequences of the extrapolations of the successive 
odd pairs of modified-ratio  approximants   $(\gamma^{bcc}(S))_n$ vs.
 $1/n^{1+\theta}$.The continuous lines are only
 guides to the eye.}
 \label{fig4}
\end{figure}

In  Fig.\ \ref{fig3}
we have  plotted   the approximant sequences $(\gamma^{bcc}(S))_n$   
 for several values of $S$  between 1/2 and $\infty$. 
The  structure of the plots  is generally consistent with
  the pattern of signs of the CCS amplitudes  already
 emerged from the study of $\beta^{bcc}_c(S)$.
For each sequence
 $(\gamma^{bcc}(S))_n $ plotted in the figure,
 we have drawn as a continuous line the  extrapolant   
based  on the 
last   odd pair of approximants, whereas a dashed 
line represents the extrapolant based on the previous odd pair.
 The small residual  
curvature of the approximant subsequences, 
 which is  due to the higher-order corrections in 
eq.(\ref{gammazinnas}),  is made manifest in Fig.\ \ref{fig3}  by the splitting
of the continuous- and the dashed-line extrapolants.
It can also be exhibited more directly by plotting  (see Fig.\ \ref{fig4})
the sequence of extrapolations of the 
 successive odd (or even) pairs of approximants. 
 
 In Table \ref{tab3}, for both lattices and for several values of the spin, 
we have reported the numerical values of the 
extrapolated exponents  with an error
corresponding to a small multiple of 
 the difference between the continuous and the dashed extrapolations.
 We have also reported the derivatives  
 of these estimates with respect to $\theta$, 
 computed at the reference 
value $\theta^{ref}=0.504$. For comparison, the same Table 
also shows the exponent estimates obtained 
from DA's, while  the results
  obtained in other recent numerical studies using MonteCarlo methods, 
 by shorter HT series, or in the RG approach, 
  will be further discussed in Sec. VIII and IX and are collected in the
 next Table\ref{tab31}.

  From Figs.\ \ref{fig3}  and \ref{fig4}, it is clear that not 
 only $a^{bcc}_{\chi}(S)$, but also the amplitudes 
 of the main subleading CCS  change sign as $S$ varies between $1$ and $2$.
This very favorable circumstance, which can also be confirmed 
numerically, for example by fitting the approximant subsequences  
to the simple asymptotic form 
$ \gamma + c_1(S)/n^{\theta} + c_2(S)/n^{1+\theta}$,  
 makes us very confident about the accuracy of the exponent estimates
 presented below. 
For each value of $S$,  a simple monotonic 
 behavior appears to have set in, since as shown in Fig. \ \ref{fig4},   
the subleading  asymptotic correction 
in eq.(\ref{gammazinnas}) generally
  works in the expected ``right'' direction. Namely, it 
tends to lower the extrapolated exponent values  
 obtained from the decreasing approximant subsequences 
 for  spin $S=1/2$ and  $S=1$, while it 
tends to raise the extrapolated values obtained from the 
 increasing subsequences for $S \ge 2$. 
Only in the $S=3/2$ case, in which both the amplitudes
 of the leading and of the subleading correction 
have the smallest absolute value,
 the approximant sequence
 is very slowly increasing and the sequence of extrapolated exponents is
 very slowly decreasing.  
 Thus we can expect that, as the 
number of available  coefficients grows large,  
the range of variation with respect to $S$  
of the extrapolated estimates of 
 $\gamma^{bcc}(S)$  will continue to shrink,  further 
improving the verification of the universality 
of the exponent with regard to the spin. More precisely,  
assuming that the general features of the 
 behavior we have described persist as the order  of the series increases,  
the successive extrapolations of the sequences $(\gamma^{bcc}(1/2))_n $  and
  $(\gamma^{bcc}(1))_n $ 
  should provide    decreasing sequences of upper bounds,   
while those of  the sequences $(\gamma^{bcc}(2))_n  $, 
$(\gamma^{bcc}(5/2))_n  $   etc. should 
give  increasing sequences of 
 lower bounds for  $\gamma$.
   
At the present order of expansion, the   exponent 
estimates obtained  by our extrapolation prescription, 
 range orderly from 1.23742, for $S=1/2$, to 1.23684 for $S=\infty$. 
 Therefore, if 
 we  now {\it assume} that universality is valid, in particular that $\gamma$ 
 is independent of $S$, the previous remarks suggest to take simply the 
 average $\gamma= 1.2371(4)$ of these extrema 
 as a first rough approximation of the exponent with an uncertainty 
corresponding to  the half-width of the range of variation.  
We can further  refine this estimate   observing that,  
for  values of the spin between $1$ and $2$, 
both the leading and the  main subleading CCS are very small,
 as it appears observing that the  exponent approximant sequences 
  have very small slopes, clearly positive for $S=1$ 
 and negative for $S=3/2$ and $S=2$.  Moreover 
the extrapolated exponent estimates are fairly insensitive to the bias 
 value of $\theta$ (for instance, we have 
$\partial \gamma^{bcc}(1)/\partial \theta \approx 0.003$,  
 $\partial \gamma^{bcc}(3/2)/\partial \theta \approx -0.0005$ and 
 $\partial \gamma^{bcc}(2)/\partial \theta \approx -0.002$ 
at $\theta =\theta^{ref}$).  Since  $(\gamma^{bcc}(1))_n  $ 
and  $(\gamma^{bcc}(2))_n  $ are very close,  
 a better estimate  for  $\gamma$  
should lie in between.  
 The extrapolation of the last odd pair of terms in the sequence 
  $(\gamma^{bcc}(1))_{n}$ yields $1.23730$,  whereas for the sequence   
 $(\gamma^{bcc}(2))_{n}$ it leads to $ 1.23699$, and therefore 
  the  rough estimate given above can be improved to $\gamma=1.23715(15)$.
 Consideration also of the sequence $(\gamma^{bcc}(3/2))_n$ suggests 
that we take $\gamma=1.2371(1)$ as our final best estimate.

A closely related procedure was proposed 
long ago in Refs.\cite{fhrb}.
These authors analyzed but never published, extensive two-variable series
 in power of $\beta$ and of a continuous Ising spin variable 
(made available by B.G. Nickel), 
using partial-differential approximants methods which indicated 
 an ``effective fixed point'' around $S=3/2$.
 Within the precision of the present calculations,
 the simple   prescription of taking  
the average of the extrapolations  of  $(\gamma^{bcc}(1))_{n}$ 
 and of $(\gamma^{bcc}(2))_{n}$, or  the extrapolation of 
 $(\gamma^{bcc}(3/2))_{n}$ as the best approximation 
 of $\gamma$, should be equally 
 (or perhaps more) effective since it also 
takes advantage of the vanishing of the main subleading correction.

Since it is not difficult to show that the leading 
CCS for any observable must also vanish for the same value 
of $S$, the same prescription can be 
 used for extracting the best value of $\nu$ from 
the  approximant  sequences $(\nu^{bcc}(S))_n$  formed by 
 the series coefficients of $\xi^2(\beta;S)$ and shown in Fig.\ \ref{fig5}. 
 The following Fig.\ \ref{fig6}  shows the sequence of the extrapolations 
of the successive odd pairs of approximants.
The slower convergence of the approximants to
 the correlation-length exponent  should not be surprising,  
simply because  $\mu_2(\beta;S)$ enters into the definition of $\xi^2$. 
 At the present orders of expansion, the behavior of the sequence of 
the extrapolated exponent values is clearly not yet  
asymptotic for $S=1/2$ and $S=1$,
 while it is much smoother and shows a slowly increasing 
trend for $3/2 \leq S \leq 3$ 
and a slowly decreasing trend for  $S> 3$.  
 Thus arguing as before, 
we can  conclude that $\nu=0.6299(2)$. 

If we bias the extrapolation procedure with
 a larger value of $\theta^{ref}$,  
  the range of variation of $\gamma^{bcc}(S)$ 
 with $S$ will be   
  expanded, to an extent that can be easily figured out from
 the data reported in Table \ref{tab3}, but the estimated central 
value of $\gamma^{bcc}$  will be  
 practically unchanged. For instance, if we adopt 
 the significantly 
 larger value $\theta^{ref}=0.54$, we find $\gamma^{bcc}(1/2)=1.23782$ and  
 $\gamma^{bcc}(\infty)=1.23661$, 
whereas  $\gamma^{bcc}(1)=1.23741$, $\gamma^{bcc}(3/2)=1.23708$ 
and $\gamma^{bcc}(2)=1.23691$  are changed 
to  a smaller extent. Averaging $(\gamma^{bcc}(1))_{n}$ 
 and  $(\gamma^{bcc}(2))_{n}$ yields $\gamma= 1.23716(25)$ and 
consideration also of  $(\gamma^{bcc}(3/2))_{n}$  
 leads to essentially the same final estimate as the one obtained 
for $\theta^{ref}=0.504$.
 In the case of the exponent $\nu$, the estimated central value is slightly
lowered to $0.6298$, well within the error bars of our previous estimate.

Using the Fisher scaling relation\cite{fisc}, 
 the exponent $\eta$
 describing the large distance falloff of the two-spin correlation-function
at the critical temperature can be estimated 
 $\eta^{bcc}=2-\gamma^{bcc}/\nu^{bcc}=0.0360(8)$.

In Fig.\ \ref{fig7} and \ref{fig8}  we have shown the 
 results of the analogous procedure of extrapolation 
for $(\gamma^{sc}(S))_n$ and $(\nu^{sc}(S))_n$.  
The main  features  are  similar to the bcc case, 
except, unfortunately, for the sign pattern of the amplitudes 
 of the leading CCS,
all of which now appear to be negative, consistently 
with the  study of $\beta^{sc}_c(S)$ by the  modified-ratio method.   
In complete analogy with the bcc case, 
for both exponents $\gamma$ and $\nu$, 
the residual curvature of the  approximant sequences 
tends to correct the extrapolations in the  
 direction expected in order that universality be realized. 
The accuracy of the  exponent estimates is notably smaller, 
 due to the anticipated slower convergence 
of the sc lattice series coefficients to their asymptotic 
form and perhaps to 
the presence of larger subleading corrections in eq.(\ref{gammazinnas}).
 The values  of the exponents 
 $\gamma$ and $\nu$  obtained from the sc series data 
are  consistent with those from the bcc data, 
 but they are affected by significantly  larger uncertainties:  we  
can roughly estimate $\gamma=1.2368(10)$ and $\nu=0.6285(20)$.

The numerical progress achieved in  this study is best 
appreciated by comparing our Figs.\ \ref{fig3} and \ref{fig5} 
 with the analogous Figs. 1 
and 2 of Ref.\cite{z81}. 
We should  first observe   that in Ref.\cite{z81}  a straightforward 
extrapolation  linear in $1/n$ was implied 
for the sequences $(\gamma^{bcc}(S))_n$ and
 $(\nu^{bcc}(S))_n$. Due to this  
choice of the plotting variable and to the  smaller extension
of the bcc series available two decades ago, 
the ``relative maximal spreads'' with respect to the spin $S$,
 of the extrapolated exponent values are
$\frac{\gamma(1/2)-\gamma(\infty)} {\gamma(1/2)+\gamma(\infty)}$
 $ \approx 2.5 \times 10^{-3}$
 and $\frac{\nu(\infty)-\nu(1/2)} {\nu(1/2)+\nu(\infty)}$
$ \approx 7.6 \times 10^{-3}$, 
respectively.
   In our study of the same lattice, 
the corresponding figures are smaller 
by nearly one order of magnitude, 
namely   the relative spread is 
now $ \approx 2.3 \times 10^{-4}$ in the case of
 $\gamma$ and $ \approx 1.4 \times 10^{-3} $ for $\nu$.
 The values of these spreads can be 
taken as rough accuracy limits for the verification of the 
universality with respect to $S$, which is 
thereby  convincingly corroborated by the new analysis.

We close this discussion with a few remarks.
 Our extension of the series to order $\beta^{25}$ has been crucial in showing 
 that, in the bcc lattice case, the asymptotic structure
 of the HT expansion coefficients  is already 
 well stabilized, since the last six or seven
 modified-ratio method  approximants  of the critical inverse temperature 
or of the critical exponents show remarkably regular trends.
  Also in the sc lattice case, there are indications   from  
 the last three or four  approximants obtained by the same method, that a 
similar trend is setting in, but clearly the convergence is  not 
as fast as for the bcc lattice.

Some numerical experimentation with model series suggests, 
 also for the exponent analysis, 
 that our error estimates
 are reasonable and  quite conservative. 

For both lattices,  the CCS amplitudes  can be estimated from 
 the slopes of the exponent approximant sequences,
 as will be further discussed in Sec. IX.   

 In conclusion, this simple  modified-ratio approach 
  confirms accurately the 
 universality of $\gamma$ and $\nu$ 
with respect to the magnitude of $S$ and to the 
 lattice structure and, conversely, {\it assuming} universality and 
using  the bcc series data, it yields
 very accurate   estimates for these exponents.

\section{ Biased modified-ratio method for the exponents} 
 In Ref.\cite{z81}, J. Zinn-Justin proposed also a
 more direct   modified-ratio procedure for biasing the exponent estimates 
with the value of $\theta$,  in order  to eliminate or strongly reduce  the 
  influence of the leading confluent corrections to scaling. 
 The prescription involves the quantities 

\begin{equation}
\bar s_n=(s_n+s_{n-1})/2
\label{seq}\end{equation}

and

\begin{equation}
b_n=\big (\frac{1}{\theta}(\bar s_n^{\theta/2}- 
\bar s_{n-2}^{\theta/2})\big )^{2/(\theta-1)}
\label{beq}\end{equation}

In terms of $b_n$ 
 the following approximant sequence can be formed 
 \begin{equation}
 (\hat \gamma(S))_n = 1 + \frac{( b_n -  b_{n-2})^2} {2( b_n + b_{n-2})}
\label{gammazinnth}\end{equation} 

 If  we  make the simplifying assumption that 
$\theta$ is exactly $1/2$, also the correction terms 
$O(1/n^{2\theta})$ will be  eliminated by this prescription,
 along with the regular correction  $O(1/n)$ and therefore 

\begin{equation} 
 (\hat \gamma(S))_n =\gamma(S)+O(\frac{1} {n^{3/2}})
\label{gammath}\end{equation}

By the remarks made at the beginning of the preceding section, these are not
 decisive improvements in the calculation of $\gamma$ and $\nu$,
and indeed,  both for the sc and  the bcc lattice, 
the results obtained by this procedure 
 are consistent with but not more accurate than 
those of our previous  analysis by  modified-ratio methods. 
See for example Figs.\ \ref{fig9} and \ref{fig10}, 
where, for convenience, we have plotted $(\hat \gamma(S))_n$ vs. 
$1/n^{2+\theta}$ rather than vs. $1/n^{1+\theta}$, because 
 the plots of the approximants appear to be more nearly linear 
 (although with somewhat large corrections) 
with respect to former than to the latter variable.

On the other hand, this  biased variant of the  modified-ratio method, 
 is more successful  in 
the analysis of the   expansions that we have computed for   $\chi_4$. 
 The sequences of biased approximants $ (\hat \gamma^{\#}_4(S))_n $  
for the exponent of $\chi^{\#}_4(\beta,S)$, 
are shown in Fig.\ \ref{fig11} for the bcc lattice  
and in Fig.\ \ref{fig12} for the sc lattice. 
In order to avoid confusing the plots,  
 we have indicated only the  extrapolations, linear 
in $1/n^{2+\theta}$, based on the last  odd pair of terms  
 $\big\{ (\hat \gamma^{\#}_4(S))_{21}, (\hat \gamma^{\#}_4(S))_{23}\big\}$ 
  in the approximant sequences. 
When the spin $S$ varies between $1/2$ and $\infty$, the extrapolated values
of $\gamma^{sc}_4(S)$ range from $4.366$ for $S=1/2$, to $4.372$ 
for $S=\infty$. 
 Similarly, in the case of the bcc lattice the values of  $\gamma^{bcc}_4(S)$ 
 vary between  $4.369$ and $4.375$.   
  We have reported in Table \ref{tab4} the
   results obtained 
by this method for several values of $S$.
From the sc lattice data we can conclude that 
$\gamma_4^{sc}= \gamma + 2\Delta=4.369(8)$    
 and from the bcc
lattice data    $\gamma_4^{bcc}= \gamma + 2\Delta= 4.372(8)$. 

The accuracy in the verification of the 
 validity of hyperscaling is often   characterized quantitatively 
 by quoting 
the value of the right-hand side of eq.(\ref{hyp}): from our estimates 
we have: $\gamma+3\nu-2\Delta = 2\gamma + 3 \nu - \gamma_4= -0.0099(160)$ 
 in the sc lattice case, and analogously $\gamma+3\nu-2\Delta=-0.0081(88)$ 
 for the bcc lattice.

 These results give strong support
 to the validity 
 of the hyperscaling relation and  
of the universality of $\gamma_4$ 
 with respect both to the lattice structure 
 and to the value of $S$.

\section{ Ratio estimates for the exponent  
of the leading confluent singularity}

 Assuming that $\theta$ is universal,
 the simplest prescription for estimating this exponent is based on  
 the series with coefficients
 \begin{equation}
q_n(S_1,S_2)=\frac{c_n(S_1)d_n(S_2)}
 {c_n(S_2)d_n(S_1)}
\label{thrat}\end{equation}
for $n>0$ and, of course, $S_1 \neq S_2$. 
Here $c_n(S)$ are the coefficients 
of the susceptibility  and $d_n(S)$
 the coefficients of the second correlation moment for spin $S$.
From eq.(\ref{asycoe}) we can observe that, for large $n$ 
\begin{equation}
q_n(S_1,S_2)=
A(S_1,S_2)[1+\frac{B(S_1,S_2)}{n^{\theta}} +O(1/n)]
\label{thratas}\end{equation}
therefore
\begin{equation}
r_n(S_1,S_2)=\frac{q_n(S_1,S_2)}
{ q_{n+2}(S_1,S_2)}= 
1+\frac{A'}{n^{\theta+1}}+O(1/n^2)
\label{thrap}\end{equation}
so that 
\begin{equation}
(\theta(S_1,S_2))_n= \frac{1} {2}\Big [n(\frac {r_n(S_1,S_2)-1}
 {r_{n+2}(S_1,S_2)-1}-1) -2\Big ] 
\label{thseq}\end{equation}
 is an approximant sequence for $\theta$.
In the bcc lattice case, if we choose $S_1=1/2$,  
$S_2> 2$, and extrapolate only the 
even  (or, equivalently, the odd) subsequences linearly in  $1/n^{1+\theta}$, 
we obtain Fig.\ \ref{fig13}. 
 The results indicate very suggestively that $\theta= 0.50^{+3}_{-1}$, 
independently of the value of $S_2$.

\section{ An  analysis of the   exponents by differential approximants} 
 The  modified-ratio methods  employed in the last sections 
have  proved   successful 
 and suggestive both for the determination of the critical 
 temperatures and for 
 the calculation of the exponents $\gamma$, $\nu$ and $\gamma_4$. 
   Let us now turn to the more traditional differential approximants (DA)
 based procedures of series
analysis after  recalling  that  their main difficulty    is the necessity of 
  some further extrapolation  
 with respect to the order of the series used, which is    
 not  straightforward, due mainly  to the lack of 
simple estimates for the finite-order corrections and to the spread
of the various DA estimates at a given order of approximation.
 This fact also hampers the assessment of the 
errors, which can be realistic 
 if not only   they reflect 
 the spread of the values of the highest-order approximants, 
but also allow for the possible residual trends.
In this respect, the  modified-ratio methods might be easier to use, as we
 have suggested in the previous sections.
We have already discussed in Sect. IV 
the DA estimates for the critical points.
 For measuring the exponents,  we   have preferred    series analyses 
 using  the inhomogeneous first- and second-order
 DA's {\it biased} with $\beta_c$ (or in some cases with $\pm \beta_c$),
 or sometimes the  {\it simplified} inhomogeneous first-order
 differential approximants  defined in \cite{bcsd}, 
 in which we have  fixed also the correction-to-scaling 
exponent $\theta$. 
The  extrapolations of the results from  the biased DA's 
and from the  simplified 
DA's   may   be  performed with a smaller uncertainty,  
 because the spread of the estimates tends to be narrower than
 for unbiased approximants.
 Moreover,  in order to understand, at least qualitatively, 
 how the estimates on a given lattice
 depend on the spin and to improve them,
 it will be sufficient   to assume that the 
 leading finite-order corrections are proportional to the amplitudes of 
the leading nonanalytic corrections to scaling(CCS).

 A simpler approach\cite{rosk,amp} similar to the 
simplified DA's consists in 
 forming the conventional Pad\'e approximants  
after subjecting the series to the biased variable change 
 $w^{\#}(S)=1-\tau^{\#}(S)^{\theta}$ 
in order to regularize the leading 
CCS.   The  results obtained either by  simplified DA's or 
by Pade' approximants in the variable  $w^{\#}(S)$ are sometimes
 numerically comparable,  but the latter
 are generally  affected by  a larger uncertainty.

 We have computed also 
the  {\it effective exponents}, introduced long ago  
in Ref.\cite{vers} and more recently reconsidered and systematically studied 
in Refs.\cite{kouv,liufi,wegried,orko},  
for the susceptibility 

\begin{equation}                       
\gamma^{\#}_{eff}(\beta;S) \equiv -
 \frac {d\log \chi^{\#}(\beta;S)} {d\log \tau^{\#}(S)}
=\gamma(S)- \theta a^{\#}_{\chi}(S)\tau^{\#}(S)^{\theta}+ O(\tau^{\#}(S)) 
\label{effg}\end{equation} 

for the correlation length

\begin{equation}                       
\nu^{\#}_{eff}(\beta;S) \equiv -(1-\tau^{\#}(S))
 \frac {d\log \xi^{\#}(\beta;S)} {d\log \tau^{\#}(S)} 
=\nu(S)- \theta a^{\#}_{\xi}(S)\tau^{\#}(S)^{\theta} + O(\tau^{\#}(S))
\label{effn}\end{equation} 

and  for the second field derivative of the  susceptibility

\begin{equation}                       
\gamma^{\#}_{4eff}(\beta;S) \equiv -
 \frac {d\log \chi_4^{\#}(\beta;S)} {d\log \tau^{\#}(S)}
=\gamma_4(S)- \theta a^{\#}_{4}(S)\tau^{\#}(S)^{\theta}+ O(\tau^{\#}(S)). 
\label{eff4}\end{equation} 

The critical exponents $\gamma$, $\nu$ 
and $\gamma_4$  are estimated by extrapolating 
  the effective exponents  
to  the critical singularity.  
Of course, 
the factor $(1-\tau(S))$ in eq.(\ref{effn})  
 is introduced only to compensate  
for the singularity  of $\frac {d\log \xi(\beta;S)} {d \log \tau(S)}$ 
at $\beta=0$ and  is unimportant at the critical point.

 It is interesting to plot the effective exponents 
  over a  wide vicinity of   
$ \beta^{\#}_c(S)$, not only to gain informations on the 
 leading correction amplitudes
 $a^{\#}_{\chi}(S)$,   $a^{\#}_{\xi}(S)$ and $a^{\#}_{4}(S)$ 
through  eqs.(\ref{effg})-(\ref{eff4}),  
  by examining whether and how fast
 they approach the critical limit  from above or from below, 
   but also simply in order to display
 the variety\cite{anis} of preasymptotic critical behaviors 
 which  can occur within  the same universality class. 
The parametrizations of the approach to the critical 
behavior, proposed within various field-theoretical 
approaches\cite{bagnu,dohm} to RG,  must 
confront also with this  phenomenology.

In Fig.\ \ref{fig14} and \ref{fig15}  we have shown the highest-order 
 simplified DA's of  
the effective exponents $\gamma^{bcc}_{eff}(S)$ and, respectively,
  $\nu^{bcc}_{eff}(S)$ 
 for spin $S=1/2,1, \ldots \infty$, over  wide ranges of inverse temperatures. 
In order to make  the curves conveniently comparable for all 
 values of the spin,  we have plotted 
 the effective exponents against the variable 
$(\tau^{bcc}(S))^{\theta}$ .
The sign of the leading CCS 
 is revealed by the slope of the plots near 
  $\tau^{bcc}(S)=0$. 

While the simplified DA's are quite sufficient to give a 
 general view of the behavior of the effective exponents, 
more  accurate results for the exponent $\gamma$ are obtained 
 extrapolating  the effective exponent expansions 
 by second-order inhomogeneous
 DA's biased with $ \beta_c$. The estimates thus obtained for $\gamma(S)$ 
range from $\gamma^{bcc}(1/2)=1.2385(6)$ to   
 $\gamma^{bcc}(\infty)=1.2367(5)$. They are reported in Table \ref{tab3}.  
 Our best DA estimate $\gamma=1.2373(4)$ is
 obtained simply  by 
averaging $\gamma^{bcc}(1)$ and $\gamma^{bcc}(2)$ and taking
 into account also the value of  $\gamma^{bcc}(3/2)$. It  
 agrees well with the  estimate by  modified-ratio methods.
The corresponding
 results for the correlation-length exponent 
 range from $\nu^{bcc}(1/2)=0.6314(20)$ to
 $\nu^{bcc}(\infty)=0.6294(5)$ and our best estimate is
$\nu=0.6301(4)$.

In the sc lattice case, the analogous (but less well converged) plots for  
$\gamma^{sc}_{eff}(\beta;S)$  and for  $\nu^{sc}_{eff}(\beta;S)$,  
obtained by  simplified DA's, are shown in Figs.\ \ref{fig16} and \ref{fig17}.
This analysis also confirms that, on the sc lattice,   
the  amplitudes of the leading CCS do not have  a  dependence on $S$  similar 
to the bcc case, but  remain  negative for all values of the spin.
The estimates of $\gamma^{sc}(S)$  and of
  $\nu^{sc}(S)$ obtained by second-order biased DA's 
 are also reported in Table\ref{tab3} .

 The simplified-DA analysis of the effective exponent
  $\gamma^{bcc}_{4eff}(S)$ of $\chi^{bcc}_4(\beta;S)$  
yields estimates of  $\gamma^{bcc}_{4}(S)$
ranging between $4.3647$ and $4.3653$. 
 It also indicates 
 that $a^{bcc}_4(1/2)$ and  $a^{bcc}_4(1)$ are negative, whereas 
$a^{bcc}_4(S)$ is positive
for $S>3/2$. 
  On the sc lattice,  the  corresponding estimates 
of $\gamma^{sc}_{4}(S)$ vary between $4.363$ and $4.373$ 
 and  $a^{sc}_4(S)>0$ for all $S$.
 We can conclude that $\gamma_4= 4.366(2)$, independently 
of the spin and the lattice, and in good agreement with the 
results of the  analysis by  modified-ratio methods. 
The accuracy in the verification of hyperscaling is now 
 slightly improved with respect to the biased  modified-ratio methods
 of sect. VI, 
 since we have $\gamma +3\nu-2\Delta=-0.0021(28)$. 

As shown in Figs.\ \ref{fig18} and \ref{fig19}, 
the pattern of signs for the confluent amplitudes of $\chi^{\#}_4(\beta;S)$
 is  consistent 
with the corresponding results for $\chi^{\#}(\beta;S)$, as it must, since 
 the ratios 
$a^{\#}_{4}(S)/a^{\#}_{\chi}(S)$  are expected to be  universal.

The exponent $\gamma^{\#}_{4}(S)$  can   also be  evaluated 
extrapolating the effective exponents 
by  inhomogeneous
 second-order DA's biased with $ \beta^{\#}_{c}(S)$.
On the bcc lattice, our results, which  appear in Table \ref{tab4}, 
range between $4.376(8)$ for $S=1/2$ and $4.3631(4)$ for  
 $S=\infty$. In particular we find $\gamma^{bcc}_{4}(1)= 4.3666(10)$,  
$\gamma^{bcc}_{4}(3/2)= 4.3638(10)$ and $\gamma^{bcc}_{4}(2)=4.3629(10) $.
 By the same arguments used in the  modified-ratio method
 analysis of the bcc lattice series,
the best value for the exponent $\gamma_{4}$ should lie 
between the estimates for $S=1$ and $S=2$. 
 This leads to our final estimate $\gamma_{4}= 4.3647(20)$, which 
 is even more accurately consistent
 with hyperscaling, since for the bcc lattice data we have  
 $2\gamma +3\nu -\gamma_4=-0.0008(28)$.

Finally, it is worth to mention briefly that 
the ratio  $\nu/\gamma$ can be determined, to a good precision,
also    studying  the log-derivative ratios  $Dlog(\chi_4)/Dlog(\chi)$
 and $Dlog(\mu_2/\beta)/Dlog(\chi)$, either by DA's biased in $\beta_c$
 or by  simplified DA's biased in $\beta_c$ and $\theta$. 
The values thus obtained from the bcc lattice expansions  (except for $S=1/2$)
 fall within the error bars of our best 
 result $\nu/\gamma=0.5092(2)$  from modified-ratio methods.
The accuracy of the estimates can be further improved by focusing on the bcc 
 lattice case
and arguing as usual that the best value of  $\nu/\gamma$  
is simply an average of the estimates
 for $S=1$ and $S=2$. 
 We thus arrive to the value $\nu/\gamma=0.5091(1)$.  
 The estimates of this ratio obtained from the sc series lie within twice 
the expected error bars for $S > 2$, but are slightly worse for smaller
 values of $S$.
In the bcc lattice case, also
 the DA estimate of $\gamma$   obtained 
 from the analysis of the ratio of the log-derivatives of $\chi'$ and $\chi$,
 whose value  at the critical point 
is $1+1/\gamma$, 
 agrees very closely with our best  results by modified-ratio methods.

We have also examined 
the term-by-term divided series 
\begin{equation} 
 Q(x;S)=  \sum_{r \geq 0} \frac {e_r(S)} {p_r(S)} x^r
\label{tbtd}\end{equation} 
 where $e_r(S)$ are the expansion coefficients of $\chi_4(\beta;S)$
 and $p_r(S)$ those of $\chi^2(\beta;S)$.
Using eq.(\ref{asycoe}), it is easily shown that 
 the auxiliary function $  Q(x;S)$ has a critical point at $x=1$ 
with exponent $ 3 \nu+1$, if hyperscaling holds.
 A  second-order biased DA analysis of the effective exponent yields 
the estimate $\nu=0.6300(4)$ independently of $S$ and 
 in complete agreement with hyperscaling.

In conclusion, we have observed that if  the sequences of  
modified-ratio method approximants 
are carefully extrapolated using as bias the value
 of $\theta$  derived by RG methods, 
the estimates of the exponents obtained  by the 
 modified-ratio method,  for all  $S$ and on both 
 lattices under consideration,   show a  
good agreement  with the results from  DA's biased only with $\beta_c$ 
or simplified DA's biased  with both $\beta_c$ and $\theta$.
 The close consistency between the critical parameter 
estimates obtained by a number of  
different procedures adds further 
confidence that the HT series have now  
reached a fairly safe extension and that 
we are not being misled by accidental apparent convergence,
so that the uncertainties of the HT  estimates 
 can be   significantly reduced.  
  The small residual  dependence of the exponent estimates 
 on the spin $S$ and on the lattice structure 
 can be confidently used to characterize 
 how accurately universality is  respected.

\section{A comparison with other exponent estimates}
The agreement  of our HT estimates
 with the values $\gamma=1.2396(13)$ and $ \nu=0.6304(13)$, 
obtained in the context of RG by Borel-summation of coupling constant
seventh-order  fixed-dimension
 perturbative expansions\cite{murnick,guida,zeq}, 
or with the values 
$\gamma=1.2380(50)$ and $ \nu=0.6305(25)$,
 obtained by Borel-summation of
 the  fifth-order
 $\epsilon-$expansion\cite{guida,zeq}, is still  reasonable. 
The values  $\gamma=1.2378(6)+0.18(g_r-1.40)$ and 
$ \nu=0.6301(5)+0.12(g_r-1.40)$ proposed in Ref.\cite{murnick} on the 
basis of a slightly different resummation of the fixed-dimension
 perturbative RG expansion, are perhaps even closer.  
At the presently available orders of HT expansion, our series
 estimates  prefer  central values for $\gamma$ and $\nu$ which are 
only  slightly lower.
It is appropriate to mention that very similar  central estimates, though
 with larger uncertainties,    
  were already obtained quite some time ago\cite{nr90,fhr,fhrb,geor,georth} 
 from bcc lattice series of order $\beta^{21}$ 
  by  the method\cite{z81,fhr} of Chen, Fisher, 
Nickel and Zinn-Justin.
 For instance, the analysis of Ref.\cite{nr90}  yielded  
$\gamma=1.237(2)$ and $ \nu=0.6300(15)$. A more recent 
study\cite{campo} of  
 HT series through $O(\beta^{20})$
for the sc lattice,  along the same lines as in Refs.\cite{nr90,fhr,fhrb}, 
indicates $\gamma=1.2371(4)$ and  $ \nu=0.63002(23)$.
 All these results are also quoted for comparison in our Table \ref{tab31}.  

 The technique\cite{z81,fhr,fhrb} of
 focusing the analysis on some particular 
model in the Ising universality class
 with negligible amplitudes of   the leading confluent 
corrections to scaling(CCS)
 was advantageously adapted also to MonteCarlo simulations 
in Refs.\cite{blh,bst}, which report 
$\gamma=1.2372(17)$ and $ \nu=0.6303(6)$. 
This procedure was further improved in   Ref.\cite{has}, in which it led to 
the estimates  $ \nu=0.6296(7)$ and $\eta= 0.0358(9)$, implying
$\gamma=1.2367(20)$.  Even lower
 central estimates of the exponents, 
namely $\gamma=1.2353(25)$ and  $ \nu=0.6294(10)$
 have  been measured in a more conventional MonteCarlo simulation of 
the spin-$1/2$ Ising model supplemented by a finite-size 
scaling analysis\cite{balle} which  allows also for 
the corrections to scaling.  It is tempting to conjecture 
 that, for $S=1/2$ on the sc lattice,  our results from the extrapolations 
of the modified-ratio approximants  and  
 the best finite-size-scaling analyses
 of the MonteCarlo simulations  on the largest accessible lattices
 are subject to errors of the same nature.
 This would explain the rather small central values of the quoted 
MonteCarlo estimates of $\gamma$ and indicate the need of   simulations of
 a significantly larger scale in order to obtain from 
spin-1/2 systems on the sc lattice
  exponent values in closer
 agreement with our bcc-series estimates.

Let us now comment briefly on the existing results for $\gamma_4$. 
We recall that the validity of the hyperscaling 
relation eq.(\ref{hyp}) for the spin-1/2 Ising model
 was  questioned\cite{bahy,bin,bfre}  on the basis of 
an analysis of 10-12 term series on the sc,
 the bcc and the fcc lattices, yielding the estimate
$\gamma+3\nu-2\Delta=0.038(12)$. This result was
 at the time interpreted as an indication of a  small, but
 clear, failure of hyperscaling. 
 As  already mentioned in the introduction, 
 the problem was convincingly settled only when the HT series for
 $\chi$ and $\xi^2$  on the bcc lattice, extended\cite{ni21}
 up to order 21, were    analyzed with careful 
allowance\cite{ni21,nr90,z81,fhr,fhrb,geor,georth,oths,adler}
 for the CCS and indicated  the insufficient accuracy of the ``classical'' HT 
 estimates  $\gamma = 1.250(3)$,  $\nu = 0.638(2)$ and $\alpha=1/8$  
generally accepted\cite{fisre,moo,ga} until then.

For general spin,  a single  study\cite{mck}
 of $\chi_4$,  performed with series $O(\beta^{13})$ on the fcc lattice, 
 can be found in the literature. The 
log-derivative  
 of the series $Q(x;S)$, defined by eq. (\ref{tbtd}), 
was examined by PA's.
 The analysis produced estimates of
   the exponent $2\Delta - \gamma$ ranging from $1.887(1)$, for $S=1/2$,
 to  $1.893(1)$, for  $S=9/2$. Of course, if hyperscaling is valid 
 $2\Delta - \gamma= 3\nu$.  Thus the final estimate 
$2\Delta - \gamma= 1.890(3)$ 
indicated the absence of hyperscaling violations of the size predicted from 
Refs.\cite{bahy,bin,bfre}, provided that
  the central value  $\nu=0.630$ suggested by  the RG, rather than the 
 ``classical'' HT estimate $\nu= 0.638(2)$, was adopted.

 The expansion of  $\chi_4$ on the sc lattice, for 
 $S=1/2$, was at the time already   
available\cite{mcke} up to order 17, but it was analyzed only later 
in Ref.\cite{g86}.
 It  yielded the estimate $\gamma_4=4.370(5)$   still confirming the validity 
of hyperscaling, provided that
 the revised values obtained from the RG in those years,  
were assumed for $\gamma$ and $\nu$.
The  series on the bcc lattice, for $S=1/2$, was  
extended to the same order  
only much later\cite{bcgn} and  its analysis 
 also confirmed the above conclusion.
 Further support of these results came also 
from various more recent MonteCarlo tests\cite{mon,tsypin,kim,kawa}.
 
It should be stressed that the finite-order effects are stronger in 
the calculation of  $\chi_4$ (and of the related quantities like $g_r$)
 than in the calculation of quantities defined 
in terms of two-spin correlations, as we have already 
remarked in Ref.\cite{bcgn}, 
and therefore that the accuracy of the results is correspondingly smaller.  
 Our comparison with previous studies, 
 shows, however, that we have achieved some improvement  not only in 
  the precision 
of the estimates  of $\gamma$ and $\nu$,
 but   mainly   of $\gamma_4$, by taking advantage of our significantly 
  extended  expansions of $\chi_4$.

\section{Estimates of critical amplitudes }

For proper reference and 
for comparison with the earlier studies,  
 we have reported in  Table \ref{tab5} a set of 
updated estimates for the critical 
amplitudes of $\chi^{\#}(\beta;S)$, of $\xi^{\#}(\beta;S)$, 
of $C^{\#}_H(\beta;S)$  
and of $\chi_4^{\#}(\beta;S)$ as defined by eqs.(\ref{conf})-(\ref{conf4}). 

We have evaluated the critical amplitude $ C^{\#}(S)$  
of $\chi^{\#}(\beta;S)$ as follows.
We have 
 adopted as a bias the value $\gamma= 1.2371$ and our estimates of 
$\beta^{\#}_c(S)$ to compute the HT series 
of  the ``effective amplitude''
\begin{equation}
C^{\#}(\beta;S)=\tau^{\gamma} \chi^{\#}(\beta;S) 
\simeq C^{\#}(S)\Big(1+ a^{\#}_{\chi}(S)
\tau^{\#}(S)^{\theta} +\ldots
+ b^{\#}_{\chi}(S)\tau^{\#}(S) + \ldots \Big)
\label{effamp}\end{equation} 
 
The  amplitude $ C^{\#}(S)$  is then 
estimated by extrapolating the effective amplitude 
 $C^{\#}(\beta;S)$ to the critical point. 
The   
extrapolation   has been performed  either by   
 first-order inhomogeneous  simplified DA's 
biased with $\beta^{\#}_c(S)$ and with $\theta$ 
in order to allow for the confluent corrections to scaling
 or, more traditionally, by 
 using second-order inhomogeneous DA's biased with $\pm \beta^{\#}_c(S)$.
 Since these two procedures yield fully consistent estimates, 
we have reported in Table \ref{tab5} only the  results obtained by the
 usual differential approximants, 
which do not need  to be biased also with $\theta$. 
 By the same procedure,  we have also 
studied the effective amplitudes for the correlation length and for $\chi_4$
 in order 
to evaluate the corresponding critical amplitudes
$ (f^{\#}(S))^2= \lim_{\tau \to 0+} \tau^{2\nu} \xi^{\#}(\beta;S)^2$ 
 and 
$C_4^{\#}(S)= -\lim_{\tau \to 0+} \tau^{\gamma+2\Delta} \chi_4^{\#}(\beta;S)$. 
 The results are reported  in the same Table.

The above mentioned difficulties in the analysis 
of the critical behavior of the specific heat, also result  into 
  larger errors of the critical amplitude $A^{\#}(S)$.   
Therefore it seems more convenient  to  compute this quantity
 from the second derivative of $G^{\#}(\beta;S)$,  which presents a sharper 
singularity.

Other estimates for some  of the mentioned
 critical amplitudes, obtained from shorter HT series  
and under slightly different biasing assumptions, 
 or by other numerical methods,
 can also be found in earlier studies\cite{georth,g86,liufi2,zfis,ruge}. 
For instance, from Ref.\cite{liufi2}  we have 
 cited in  Table \ref{tab5} the estimates of  $A^{sc}(1/2)$,  $C^{sc}(1/2)$ 
and $f^{sc}(1/2)$, obtained 
from series $O(\beta^{17})$, $O(\beta^{19})$ and $O(\beta^{12})$
 respectively,  under the assumptions $\alpha=0.104$,
 $\gamma=1.237$, $\nu=0.6325$  and
$\beta_c^{sc}(1/2)=0.221620$. From the same reference, 
we have also reported  the 
estimates of $A^{bcc}(1/2)$ obtained from 
a series $O(\beta^{17})$, and 
of $C^{bcc}(1/2)$ and $f^{bcc}(1/2)$, obtained from 
series  $O(\beta^{21})$, 
 by assuming $\beta^{bcc}_c(1/2)=0.157362$ 
and the same values as above for $\alpha$, $\gamma$ and $\nu$.
 
Under various assumptions on the value of $\alpha$, estimates of 
  $A^{sc}(1/2)$ were derived in Ref.\cite{haspi} from a  simulation
 in which the energy and the specific heat were measured. By straightforward
 interpolation, we can conclude that, for $\alpha=0.11$, these data would imply
 the estimate  $A^{sc}(1/2)=1.368(7)$, in reasonable agreement with ours.
 
From  Ref.\cite{zfis}, we have  quoted estimates 
of $C^{sc}_4(1/2)$ and of $C^{bcc}_4(1/2)$ obtained from series 
$O(\beta^{17})$ and $O(\beta^{13})$, respectively,  
assuming $\gamma_4=4.375$ and   
 $\beta_c^{sc}(1/2)=0.221630(16)$,  $\beta_c^{bcc}(1/2)=0.157368(7)$.

 In the same Table, we have also reported  
 the estimates\cite{georth}
 from series $O(\beta^{21})$  for $C^{bcc }(1/2)$,
 $C^{bcc }(1)$ and $C^{bcc }(2)$    assuming  $\gamma=1.237$
  and  the estimates of 
 $f^{bcc }(1/2)$, $f^{bcc }(1)$ and  
$f^{bcc }(2)$,  
 obtained assuming   $\nu=0.6297$, $\nu=0.6298$, 
  $\nu=0.6300$ respectively,   together with the values of 
$\beta^{bcc}_c(S)$  quoted in the same Reference and reported 
in Table \ref{tab2}.

 In Ref.\cite{ruge} the values of $C^{sc}(1/2)$ and of $f^{sc}(1/2)$ have been
 computed by a MonteCarlo method assuming  $\gamma=1.237$, $\nu=0.628$ and 
$\beta_c^{sc}(1/2)=0.22165$. 
 
This brief review of some existing results shows how  sensitively 
the estimates of the critical amplitudes depend on  
the bias values chosen for the critical exponents and  temperatures, and, of 
 course, on the length of the series.
  If we  also allow properly  for these sources of 
uncertainty, which generally are not included 
in the  error bars quoted in the literature, many of the cited  estimates 
can be considered essentially compatible among themselves and
with ours.

\section{Estimates    of the renormalized coupling }
 The value of the hyper-universal renormalized coupling constant $g_r$
 can be obtained  from our estimates of the critical amplitudes.
  Alternatively, without biasing the computation also with the critical 
exponents, $g_r$ can be computed, with a smaller uncertainty 
 by extrapolating to critical point  the expansion of the 
auxiliary function
 \begin{equation}
 y^{\#}(\beta;S)= (g_r^{\#}(\beta;S))^{-2/3}
\label{yg}\end{equation}
 or of 
 the function
 \begin{equation}
z^{\#}(\beta;S)=(\beta/\beta^{\#}_c(S))^{3/2} g_r^{\#}(\beta;S).
\label{zg}\end{equation}
 Unlike the effective coupling $g_r^{\#}(\beta;S)$,
 both  $y^{\#}(\beta;S)$  and $z^{\#}(\beta;S)$ are regular analytic at 
$\beta=0$, so that  they
 can be expanded in powers of $\beta$ and extrapolated to the critical 
point  by Pad\`e approximants, DA's or simplified DA's. 
Due to the finite extension of the series, the numerical 
estimates of $g_r$ derived from eq.(\ref{yg}) or from eq.(\ref{zg})    
 are of course  (very) slightly different.
 In order to allow  for the expected  
 leading confluent corrections to scaling(CCS), 
in  our calculations we have used first- and  
 second-order DA's biased in $ \beta^{\#}_c(S)$  or  simplified DA's 
 biased with $\beta^{\#}_c(S)$ and with  the 
confluent exponent $\theta$.
In Fig.\ \ref{fig20} we have plotted vs.  $\tau^{bcc}(S)$ the effective 
coupling $g_r^{bcc}(\beta;S)$ as obtained from 
the function $z^{bcc}(\beta;S)$  for various  values of the spin $S$. 
The curves, computed in the simplest way   by simplified DA's, 
   show the strong influence of the CCS nearby the critical point and 
 indicate that $g_r^{bcc}(S)$ 
is  independent of $S$ within a very good approximation. 
Comparison with Fig.\ \ref{fig21},
 which shows the effective coupling  $g_r^{sc}(\beta;S)$  
plotted vs. $\tau^{sc}(S)$, similarly indicates that
  the renormalized coupling $g_r^{sc} \approx g_r^{bcc} \approx 1.41 $, 
is independent not only of $S$, but also of  the lattice structure.
 Using eq.(\ref{confgg}), we can  infer from  Fig.\ \ref{fig20} 
that the amplitudes of the CCS are generally large and, more precisely, 
that $a_g^{bcc}(S) >0$ for $S=1/2$ and $ 1$,
 whereas  $a_g^{bcc}(S) <0$ for $S>2$.
Analogously, from Fig.\ \ref{fig21} we can   conclude 
that $a_g^{sc}(S) >0$ for all $S$. 
These qualitative conclusions are consistent with 
 RG estimates\cite{aha,chre} in the fixed-dimension approach indicating that 
 $a_g/a_{\chi}$ lies in a  range from  $\approx -3.$ to $\approx -2.$.

In order to reach higher precision in the calculation of $g_r$,  
 we have preferred to use first- and second-order DA's biased 
with $\beta_c$.  
In the bcc lattice case, we notice that, for $S=1/2$ and $1$, 
 approximants which use an increasing number of  
 coefficients show a residual slowly 
decreasing trend, while, for $S \geq 2$, 
 they show an increasing trend. We shall 
indicate by asymmetric uncertainties 
 these features of the approximant sequences. Again  
arguing like for the critical exponents
 $\gamma$, $\nu$, and $\gamma_4$, 
we can expect that the most reliable estimate of $g_r$ 
 obtained from the bcc lattice series  will be 
 nearly equal to the average of $g^{bcc}_r(1)$ and $g^{bcc}_r(2)$
or to the value $g^{bcc}_r(3/2)$.
Thus we conclude   that $g_r=1.404(3)$.   
The central value of our updated estimate is
 slightly lower than our previous result\cite{bcgn} $g_r=1.407(6)$, based
on  a series $O(\beta^{17})$ for the bcc lattice 
in the  $S=1/2$ case, in which the convergence is slowest. 
 However, our  revised result is slightly closer  
 to the value $g_r=1.400$  advocated  by D.B.Murray 
and B.G.Nickel\cite{murnick} in the context 
of the RG fixed-dimension approach,
 and is compatible  with the more recent HT 
result\cite{campo} $g_r=1.402(2) $ 
obtained by the method of Refs.\cite{z81,fhr} 
 from sc lattice series extending to order $\beta^{18}$.     
 Our numerical estimates of $g_r^{\#}(S)$,  for several
 values of $S$ and on both lattices, are reported in Table \ref{tab6}.
 Notice that in the bcc case the DA estimates are larger
 or smaller (in the sc case 
 generally larger) than the expected best value,  
consistently with the signs of 
 the CCS amplitudes.
In the same Table  we have  quoted for comparison   
also some recent MonteCarlo  estimates\cite{mon,kim} of $g_r$,
 as well as other results from HT and RG methods.

\section{Estimates    of $R^+_{\xi}$ }

 The combination of critical amplitudes 
$R^+_{\xi}$,  defined by eq. (\ref{confrx}),  can be computed  either  from 
the estimates of the critical amplitudes $A^{\#}$, $f^{\#}$ and of the
 exponent $\alpha=2-3\nu$, or, more directly, by extrapolating 
 to the critical point  the  expansion of the auxiliary function 
\begin{equation}
F(\beta;S)=
\frac{-q\nu^3\beta_c(S)} {2v\beta}
 \frac{d^2G(\beta;S)}{d\beta^2}
\Big(\frac{\beta^{3/2}}{\beta^{3/2}_c(S)} 
\frac{d(1/ \xi(\beta;S))}{ d\beta}\Big)^{-3}
\label{rxipiu}\end{equation}
since the validity of the hyperscaling relation 
eq.(\ref{hyp2ma}) implies  
$F(\beta_c(S)-0;S)=(R^+_{\xi})^{3}+O(1/n^{\theta})$.
 We have assumed $\nu=0.6299$ and  evaluated $F(\beta;S)$ 
by  first-order differential approximants biased with 
$\pm \beta^{\#}_c(S)$.
The estimates of $R^+_{\xi}$ obtained by this prescription are shown  
in Table \ref{tab7}.  Within a fair approximation, they are
  independent 
 of $S$ and of the lattice structure and  compatible with 
the estimates obtained combining the amplitudes reported in Table\ref{tab5} . 
 
Our final estimate is 
$R^+_{\xi}=0.2668(5)$. 
This result is slightly smaller than that the value $R^+_{\xi}=0.272(4)$
 reported in our previous study\cite{bcsd} of the single 
$S=1/2$ case employing   shorter series.

In  Table \ref{tab7} we have also shown  values of $R^+_{\xi}$
 obtained via RG, either by the 
fixed-dimension perturbative expansion to fifth order\cite{berxi}   
 or by the $\epsilon$-expansion to second order\cite{bereps}.  
 We have also quoted the estimate $R^+_{\xi}=0.270(4)$
 that would be obtained  
 from the MonteCarlo measures of Ref.\cite{haspi} 
assuming $\alpha=0.11$ and the 
recent HT result $R^+_{\xi}=0.2659(4)$ taken
 from Ref.\cite{campo} as a representative of various
 nearly equal central estimates from  studies 
\cite{bcsd,campo,liufi2} performed at different times, 
with different techniques, under different assumptions 
on the values of $\nu$ and $\alpha$
 and using series of different extensions.
 The discrepancy from our estimate should probably 
 be taken as an indication of the remaining 
difficulty of  accurately 
 evaluating  the specific heat amplitude.

Finally, it is worth while to quote  
two  recent  very accurate measurements on binary mixtures: 
 $R^+_{\xi}=0.284(18)$, performed in a  microgravity 
experiment\cite{migr}, and $R^+_{\xi}=0.265(6)$ 
 obtained in a conventional environment\cite{now}.

\section{Estimating the   ratios of  confluent-singularity amplitudes}

 From the extended series presented here, we have also tried 
 to evaluate
 the universal ratio $a_{\xi}(S)/a_{\chi}(S)$. 
We recall that, for  the bcc lattice, our analysis  of $\chi$ and $\xi$ 
 by modified-ratio methods had  shown that, as the spin 
$S$ varies between $1$ and $2$, 
the leading correction amplitudes $a_{\chi}(S)$ and  $a_{\xi}(S)$ 
 vary from small negative values to small 
positive  values, whereas  
 in the sc lattice case  no
  change of the sign of the confluent amplitudes is observed.
 As we have emphasized, 
 some knowledge of these amplitudes is necessary to 
 understand how   the various numerical estimates obtained 
for each value of $S$ approach the true values of the universal 
 critical parameters.   
A simple prescription  to compute accurately
 the universal quantities
 consists in   using 
 series on the bcc lattice with spins between $1$ and $2$, 
  for which the amplitudes of 
the leading confluent corrections to scaling(CCS) are very small. 
 Conversely,   the numerical methods to evaluate the amplitudes and the 
exponent of  the leading CCS,
can be expected to work with fair  accuracy
 only when the confluent  corrections  are {\it not} too small.
For $S=1/2$ the size of the leading CCS is largest, 
but unfortunately also the higher-order corrections seem to be 
still important, as shown by the steep 
behavior of the extrapolated sequences in Fig.\ \ref{fig4}.
 Therefore,  the most reliable results are likely to come
 from the bcc series for $S>2$.

We have obtained reasonably accurate  estimates of 
the CCS amplitudes for $\chi$  simply by fitting  
 the asymptotic form $ \gamma + c_1(S)/n^{\theta} + c_2(S)/n^{1+\theta}$, 
suggested by eq.(\ref{gammazinnas}), to the exponent 
 approximant sequences eq.(\ref{gammazinn}), under the assumptions that 
 $\gamma=1.2371$  and $\theta=0.504$. 
A similar procedure can be repeated 
 in the case of $\xi^2$, assuming $\nu=0.6299$.
  The estimates 
 of the CCS amplitudes and of  the universal ratios  
$a^{bcc}_{\xi}(S)/a^{bcc}_{\chi}(S)$ thus obtained,  
    are shown in Table \ref{tab8}. 
The values of the amplitudes $a^{bcc}_{\chi}(S)$  obtained 
 by this procedure are generally consistent   
within  a few percent, and those of $a^{bcc}_{\xi}(S)$  within 
$\approx 10-20\%$ in the worst cases, 
with those evaluated by simplified DA's of the 
log-derivatives  of $\chi$ and $\xi$, biased with 
$\beta_c$ and $\theta$. Moreover, for $S>2$, the estimates of 
$a^{bcc}_{\chi}(S)$ obtained from the modified-ratio 
approximants for $\gamma$
 are consistent within a few percent with those obtained from the 
 corresponding  approximants for $\beta_c$. As expected, for smaller values
of $S$, we have consistency only within $20-40\%$ because
 the rate of convergence of the series is slower
  and/or  the subleading confluent corrections are more important. 
  
 The ratios $a^{bcc}_{\xi}(S)/a^{bcc}_{\chi}(S)$ appear 
to be approximately independent 
of the spin $S$, as they should, and suggest the final estimate
 $a_{\xi}/a_{\chi}=0.76(6)$. The error  includes also  the 
uncertainties  of the bias parameters $\gamma$, $\nu$  and $\theta$.
In the sc lattice case, the same analysis leads 
  to amplitude ratios which  show larger uncertainties, but agree 
 within the errors with the bcc results. 

 For $S=1/2$, the series $O(\beta^{21})$ of Ref.\cite{nr90} yielded
  the estimates (without indication of error)
  $a^{bcc}_{\xi}(1/2)= -0.11$, which is   $10\%$ larger than  ours, 
 and   $a^{bcc}_{\chi}=-0.13$,  which agrees closely with ours. 

Using the same series,  Ref.\cite{georth} obtained estimates of  
$a^{bcc}_{\chi}$ and $a^{bcc}_{\xi}$  for $S=1/2,1,2$, also quoted 
in Table \ref{tab8} and in  good agreement with ours.
 
 Our central estimate of the ratio $a_{\xi}/a_{\chi}$ is
  somewhat smaller than our previous estimate\cite{bc23}
$a_{\xi}/a_{\chi}=0.87(6)$ based on shorter $S=1/2$ series, than
  the old HT estimates $a_{\xi}/a_{\chi}=0.83(5)$ of 
 Ref.\cite{geor} and   
 $a_{\xi}/a_{\chi}=0.85$ of Refs.\cite{nr90,georth} 
(reported without indication of error), 
but it is larger than the HT estimate $0.71(7)$ of Ref.\cite{z81},
 and the earlier estimate\cite{ferer}
 $a_{\xi}/a_{\chi}=0.70(2)$, 
based on the fcc series $O(\beta^{12})$ for general 
spin tabulated in Ref.\cite{camp,camp2}.  
 We should also mention the estimate $a_{\xi}/a_{\chi}=0.65(5)$ 
   obtained by  RG  in the perturbative fixed-dimension approach at  
sixth order\cite{babe}. 
 The     $\epsilon$-expansion scheme (extending to second order) 
  yielded\cite{chre} the estimate $a_{\xi}/a_{\chi}=0.65$.

Finally, let us note that our results  confirm 
the  observations of
Refs.\cite{nr90,fhr,fhrb,geor,bcsd} 
and the  arguments presented in Ref.\cite{liufi}  
 that the  amplitudes of the leading CCS 
 have a negative sign,  both for the susceptibility and for the 
correlation length, in the case of 
the spin-$1/2$ Ising model, on the sc and the bcc  lattices.

\section{ Conclusions}

For the Ising models of general spin $S$,   
on the sc and the bcc lattices, 
 we have produced   extended HT expansions  
of the nearest-neighbor correlation function, 
of the susceptibility, of the second correlation moment
 and  of the second field derivative of the susceptibility.

Our procedure of series analysis   
 differs somewhat from the most traditional ones, but leads to 
completely consistent
conclusions. At least for the models studied here,
 we are also confident that it yields  very accurate
  direct estimates of the various critical 
parameters. Our  updated results: $\gamma=1.2371(1)$, $\nu=0.6299(2)$,
 $\gamma_4=4.3647(20)$, $g_r=1.404(3)$ and  $R^+_{\xi}=0.2668(5)$ 
are  in  good agreement 
 with the latest calculations by  other approximate  
methods, including  the perturbative field theoretic RG approaches.
At the same time, our new series data   
have proven  to be sufficiently rich that we can obtain fairly tight 
checks of the  conventional expectations about 
 hyperscaling and universality, with regard 
both to the spin $S$ and to the lattice structure. 
 
\acknowledgments 
This work has been partially supported by the 
Ministry of Education, University and Research.

\newpage

\begin{table}
\squeezetable
\caption{ Orders of  high-temperature expansions, 
 published (or obtainable from published data)
 before our  work, for the nearest-neighbor correlation function 
$G(\beta;S)$, 
for the susceptibility $\chi(\beta;S)$, for the second moment of the 
correlation function $\mu_2(\beta;S)$  and for 
the second field-derivative of the susceptibility $\chi_4(\beta;S)$, 
  in the case of  the  Ising   models with general spin $S$.}
\label{tab1}
\begin{tabular}{lccc}
  Observable & Lattice; Order[Ref.] & Lattice; Order[Ref.] 
& Lattice; Order[Ref.]     \\
\tableline

$\chi(\beta;S)$ &sc; 15 \cite{sack}&bcc; 21 \cite{nr90}&fcc; 14
 \cite{mckest}  \\

$\mu_2(\beta;S)$ & sc; 15 \cite{sack}&bcc; 21 \cite{nr90}
&fcc; 14 \cite{mckest} \\

$\chi_4(\beta;S)$ & sc; 14 \cite{lw}& bcc; 10 \cite{bin}
&fcc; 13 \cite{mck} \\

$G(\beta;S)$ && &fcc; 14 \cite{mckest}  \\
\end{tabular}
\end{table}

\begin{table}
\squeezetable
\caption{ Estimates of the critical inverse temperatures 
 for the spin-$S$ Ising   models
 on the sc  and the bcc lattices, obtained in this work
 by the  modified-ratio  method 
 biased with the leading correction-to-scaling exponent $\theta$.
The sensitivity of the estimates to the bias value of $\theta$
is characterized by the derivatives of $\beta_c$.
For comparison, we have also reported some results obtained  
 by simulation methods or from the analysis of shorter
 series in the recent literature.}
\label{tab2}
\begin{tabular}{lccccccc}
 &  S=1/2 & S=1 & S=3/2 &  S=2&  S=5/2&S=3&  S=$\infty$ \\
\tableline
$\beta_c^{sc}(S)$ &0.221655(2)&0.312867(2)
&0.368657(2)&0.406352(3)&0.433532(3)&0.454060(3)&0.601271(3) \\
$\partial\beta_c^{sc}(S)/ \partial \theta$ &$7\cdot 10^{-6}$
&$4\cdot 10^{-6}$&$4\cdot 10^{-6}$&$4\cdot 10^{-6}$&
$6\cdot 10^{-6}$&$6\cdot 10^{-6}$&$6\cdot 10^{-6}$\\
\tableline
$\beta_c^{bcc}(S)$  &0.1573725(10)&0.224656(1) 
&0.265641(1) &0.293255(2)&0.313130(2)&0.328119(2)&0.435085(3)\\
$\partial\beta_c^{bcc}(S)/ \partial \theta$ &$3\cdot 10^{-6}$
&$ 10^{-8}$&$-3\cdot 10^{-6}$&$-6\cdot 10^{-6}$&
$-6\cdot 10^{-6}$&$-6\cdot 10^{-6}$&$-6\cdot 10^{-6}$\\
\tableline
$\beta_c^{sc}(S)$\cite{bst}  &0.2216546(10)&&&&&&\\
\tableline
$\beta_c^{sc}(S)$\cite{ito}  &0.2216595(15)&&&&&&\\
\tableline
$\beta_c^{sc}(S)$\cite{tama}  &0.221655(1)&&&&&&\\
\tableline
$\beta_c^{bcc}(S)$\cite{z81} &0.157374(7)&0.224657(4) 
& &0.293258(6)& & & 0.435089(11) \\
\tableline
$\beta_c^{bcc}(S)$\cite{georth} &0.157373&0.224654 
& &0.293255&& &  \\
\end{tabular}
\end{table}

\begin{table}
\squeezetable
\caption{Estimates of the critical exponents  $\gamma$ and $\nu$
 for the spin-$S$ Ising   model series on the sc  and the bcc lattices 
 obtained from  
 extrapolation of the  modified-ratio (MR) approximant sequences, 
(defined by eq.(\ref{gammazinn})). 
 We have also reported the  estimates obtained
  in this paper by  second-order inhomogeneous differential
 approximants (DA) biased with $\beta_c$.}

\label{tab3}
\begin{tabular}{cccccccc}
Exponent &  S=1/2 & S=1 & S=3/2 &  S=2&  S=5/2& S=3&  S=$\infty$ \\
\tableline
$\gamma^{sc}(S)$(MR)&1.2375(6) &1.2378(7)&1.2371(8)&1.2367(10)
 &1.2364(10)&1.2363(10)&1.2359(15) \\
$\partial \gamma^{sc}(S)/\partial \theta$&0.016&0.009&0.007&0.006&0.006
&0.006&0.006\\
\tableline
\tableline
$\gamma^{bcc}(S)$(MR)& 1.23742(20) &1.23730(16)&1.23710(3) &1.23699(10) 
&1.23694(10)&1.23691(10) &1.23685(15) \\
$\partial \gamma^{bcc}(S)/\partial \theta$&0.012&0.003&-0.0005&-0.002
&-0.003&-0.004&-0.006\\
\tableline
\tableline
$\gamma^{sc}(S)$(DA)&1.238(2) &1.239(2)&1.240(3)
&1.239(2)&1.239(2)&1.239(2)&1.238(2) \\
\tableline
$\gamma^{bcc}(S)$(DA) &1.2378(8)&1.2385(15) &1.2370(4) 
&1.2365(4)&1.2366(4)&1.2366(4) &1.2367(4) \\
\tableline
\tableline
$\nu^{sc}(S)$(MR) &0.6277(30) &0.6279(30) &0.6283(20) &0.6285(20)
 &0.6286(20) &0.6286(20)&0.6288(20) \\
$\partial \nu^{sc}(S)/\partial \theta$ &0.021
&0.014 & 0.010&0.0095& 0.009&0.008 &0.006\\
\tableline
\tableline
$\nu^{bcc}(S)$(MR) &0.6283(20)&0.6294(8) &0.6297(6) &0.6298(6)
&0.6299(6) &.6299(6)& 0.6301(5) \\
$\partial \nu^{bcc}(S)/\partial \theta$ &0.006&0.001&-0.001&-0.0015
&-0.002&-0.0022&-0.003\\
\tableline
\tableline
$\nu^{sc}(S)$(DA) &0.632(2) &0.631(1)&0.631(1)&0.631(1)&0.631(1)&
0.631(1)&0.631(1) \\
\tableline
$\nu^{bcc}(S)$(DA) &0.6308(10)&0.631(1) &0.6299(4) &0.6295(4)&0.6296(4) &
.6295(4)& 0.6294(6) \\
\end{tabular}
\end{table}

\begin{table}
\squeezetable
\caption{ Estimates of the exponents $\gamma$ and $\nu$
 obtained in the recent literature by various kinds of analyses of shorter
high-temperature series, by MonteCarlo methods or by renormalization-group
 methods. 
The estimates marked with an asterisk 
are obtained  by procedures implying or assuming universality.}
\label{tab31}
\begin{tabular}{ccccc}
Exponent & HT & MonteCarlo & RG $\epsilon$-exp. &  RG fixed-D exp. \\
\tableline
&1.237(2)*\cite{nr90}& &&\\
&1.2385(25)*\cite{z81}&& &\\
&1.2385(15)*\cite{fhr}&1.2367(11)\cite{has}&&1.2378(6)*\cite{murnick}\\
$\gamma$&1.2395(4)*\cite{fhrb}&1.237(2)*\cite{blh}
&1.2380(50)*\cite{guida}&1.2396(13)*\cite{guida}\\
&1.2378(6)\cite{geor}&1.2372(17)*\cite{bst}&&\\
&1.2373(1)\cite{georth}&&& \\       
&1.2371(4)\cite{campo}&&&\\
\tableline
   &0.6300(15)*\cite{nr90}&&&\\
  &0.6305(15)*\cite{z81}&0.6296(7)\cite{has}&&0.6301(5)*\cite{murnick}\\
$\nu$ &0.632(1)*\cite{fhrb}&0.6301(8)*\cite{blh}&0.6305(25)*\cite{guida} 
&0.6304(13)*\cite{guida}\\ 
     &0.6311(3)*\cite{geor}&0.6303(6)*\cite{bst}&&\\ 
    &0.6300(2)\cite{georth}&0.6294(10)\cite{balle}&&\\
    &0.63002(23)*\cite{campo} &&&\\
\end{tabular}
\end{table}

\begin{table}
\squeezetable
\caption{Estimates of the critical exponent $\gamma_4$ 
 from the spin-$S$ Ising   model high-temperature series
 on the sc  and the bcc lattices obtained  in this work 
by the  modified-ratio (MR) method
 {\it directly} biased with the leading correction-to-scaling
 exponent $\theta$ 
following eq.(\ref{gammazinnth}) or by second-order 
differential approximants biased with $ \beta_c$.
For comparison, we have also reported some results obtained  
 by simulation methods or from shorter
 series in the recent literature.} 
\label{tab4}
\begin{tabular}{cccccccc}
Exponent &  S=1/2 & S=1 & S=3/2 &  S=2&  S=5/2& S=3&  S=$\infty$ \\
\tableline
$ \gamma_4^{sc}(S)$ (DA)&4.372(8)&4.368(8)
&4.369(8)&4.369(8)&4.368(8)& 
4.367(4)&4.366(4) \\
\tableline
\tableline
$\hat \gamma_4^{sc}(S)$  (MR)&4.3703(12)&4.3662(16)&4.3671(9)
&4.3683(9)&4.3691(9)&4.3697(9)
 &4.3719(2) \\
$\partial\hat \gamma_4^{sc}(S)/\partial \theta$ &0.027&0.023&
0.015&0.009&0.004&0.002&-0.007\\
\tableline
\tableline
$ \gamma_4^{bcc}(S)$ (DA) &4.376(20)& 4.3666(10) &4.3638(10)
 &4.3629(10)&4.3632(10)& 4.3631(10)&4.3631(10)\\
\tableline
\tableline
$\hat \gamma_4^{bcc}(S)$ (MR) &4.3696(3)&4.3690(11) &4.3702(12)
&4.3714(12)&4.3722&4.3729(12) &4.3749(10) \\
$\partial \hat \gamma_4^{bcc}(S)/\partial \theta$ &0.02&-0.008&-0.014&
-0.023&-0.028&-0.032&-0.043\\
\tableline
\tableline
$\gamma_4^{sc}(S)$\cite{bcgn}&4.361(8)&&&&&&\\
\tableline
$\gamma_4^{bcc}(S)$\cite{bcgn}&4.366(6)&&&&&&\\
\tableline
$\gamma_4^{bcc}(S)$\cite{g86}&4.370(14)&&&&&&\\
\end{tabular}
\end{table}

\begin{table}
\squeezetable
\caption{  Estimates  
 of the critical amplitudes $A^{\#}(S)$ of the specific 
 heat $C_H^{\#}(\beta,S)$,  $C^{\#}(S)$ 
of the susceptibility $\chi^{\#}(\beta;S)$, 
 $f^{\#}(S)$ of the correlation length $\xi^{\#}(\beta;S)$ and 
 $C^{\#}_4(S)$ of $\chi^{\#}_4(\beta;S)$ for the spin-$S$ 
Ising   models  on the sc  and the bcc lattices. 
They are obtained by differential approximants 
biased with the critical inverse temperatures
 reported in Table II and with the critical exponents estimated in this work. 
 For comparison, we have also reported
 a few estimates obtained by simulation methods or from shorter
 series in the recent literature. For some of them,
 no indication of error is available. }
\label{tab5}
\begin{tabular}{lccccccc}
 &  S=1/2 & S=1 & S=3/2 &  S=2&  S=5/2&S=3&  S=$\infty$ \\
\tableline
$A^{sc}(S)$ &1.34(1)&1.80(3)
&2.00(2)&2.09(1)&2.15(2)&2.18(2)&2.28(1) \\
\tableline
$\partial A^{sc}(S)/ \partial \beta^{sc}_c(S)$&640.&980.&170. 
&550.&300. &750.& 440. \\
\tableline
$\partial A^{sc}(S)/ \partial \alpha $&-20.&-26.&-28.&-32.&-32.&-32.&-33.\\
\tableline
\tableline
$A^{bcc}(S)$ &1.302(6)&1.732(6)
&1.911(6)&2.000(6)&2.051(6)&2.082(6)&2.171(6)\\
\tableline
$\partial A^{bcc}(S)/ \partial \beta^{bcc}_c(S)$&800.&1000.&950.&
900.&1200.&870.&830.\\
\tableline
$\partial A^{bcc}(S)/ \partial \alpha$&-19.&-26.&-29.&-30.&-31.&-32.&-33.\\
\tableline
\tableline
$A^{sc}(S)$\cite{liufi2}&1.464(7)&  & && &&  \\
\tableline
$A^{sc}(S)$\cite{haspi}&1.368(7)&  & && &&  \\
\tableline
$A^{bcc}(S)$\cite{liufi2}&1.431(9)&  & && &&  \\
\tableline
\tableline
$C^{sc}(S)$&1.127(3)&0.682(1)
&0.545(1)&0.482(1)&0.443(1)&0.4184(6)&0.3073(2) \\
\tableline
$\partial C^{sc}(S)/ \partial \beta^{sc}_c(S)$&1900.&1800.&1900.&
1000.&600. &220.&150.  \\
\tableline
$\partial C^{sc}(S)/ \partial \gamma $&-9.&-8.&-7.
&-7.&-5.5 &-5.& -3. \\
\tableline
\tableline
$C^{bcc}(S)$ &1.042(1)&0.622(1)
&0.4967(3)&0.4379(3)&0.4045(4)&0.3826(4)&0.2817(4) \\
\tableline
$\partial C^{bcc}(S)/ \partial \beta^{bcc}_c(S)$&3900.&-1300.&170. 
&750.&500. &480.& 310. \\
\tableline
$\partial C^{bcc}(S)/ \partial \gamma $
&-9.&-5.&-4.5&-3.6&-3.1 &-3.&-2.2\\
\tableline
\tableline
$C^{sc}(S)$\cite{liufi2} &1.1025(10)&  & && &&  \\
\tableline
$C^{sc}(S)$\cite{ruge} &1.093(13)&  & && &&  \\
\tableline
$C^{bcc}(S)$\cite{georth} &1.026& 0.620& & 0.4346&&\\
\tableline
$C^{bcc}(S)$\cite{liufi2} &1.0312(10)&  & && &&  \\
\tableline
\tableline
$f^{sc}(S)$  &0.506(1)&0.458(1) 
&0.443(1) &0.436(1)&0.432(1) & 0.430(1)&0.423(1) \\
\tableline
$\partial f^{sc}(S)/ \partial \beta^{sc}_c(S)$&290.&280.&200.
&180.&400.&120.&120.\\
\tableline
$\partial f^{sc}(S)/ \partial \nu $&-4.&-4.&-3.5&-4.&-4.
 &-3.& -3.5 \\
\tableline
\tableline
$f^{bcc}(S)$ &0.4686(4)&0.4262(8) 
&0.4112(4) &0.4047(4)&0.4013(4) &0.3992(4)& 0.3937(4) \\
\tableline
$\partial f^{bcc}(S)/ \partial \beta^{bcc}_c(S)$&500.&450. 
 &500. &230.&200. &170.&1700.   \\
\tableline
$\partial f^{bcc}(S)/ \partial \nu $&-4.&-5.&-3.&-3.&-3. &-3.&-3. \\
\tableline
\tableline
$f^{sc}(S)$\cite{liufi2} &$0.496(4)$&  & && &&  \\
\tableline
$f^{sc}(S)$\cite{ruge} &$0.501(2)$&  & && &&  \\
\tableline
$f^{sc}(S)$\cite{georth} &$0.5192$&  & && &&  \\
\tableline
$f^{bcc}(S)$\cite{liufi2} &0.4590(1)&  & && &&  \\
\tableline
$f^{bcc}(S)$\cite{georth} &0.46821& 0.42605& & 0.4038&&\\
\tableline
\tableline
$C_4^{sc}(S)$ &3.87(1)&1.05(1)&0.606(1)&0.456(2)&0.375(1)&
0.327(1)& 0.169(1) \\
\tableline
$\partial C_4^{sc}(S)/ \partial \beta^{sc}_c(S)$&5000.&3200.
&2900. &2700. &800. &-1300.&350.  \\
\tableline
$\partial C_4^{sc}(S)/ \partial \gamma_4 $&-20.&-9.&-7.&-6.
&-5. &-1.5& -1.5 \\
\tableline
\tableline
$C_4^{bcc}(S)$ &3.410(8)&0.912(3)  
&0.523(1) &0.3884(6)&0.3230(5) &0.2847(5)&0.1478(5)\\
\tableline
$\partial C_4^{bcc}(S)/ \partial \beta^{bcc}_c(S)$&15000.&8000.  &460.
 &630. &470. &450.&190.  \\
\tableline
$\partial C_4^{bcc}(S)/ \partial \gamma_4 $&-24.&-9.5&-3.&-2.5&-2.&-2.&-1.\\
\tableline
\tableline
$C_4^{sc}(S)$\cite{g86} &3.70(3)&  & && &&  \\
\tableline
$C_4^{sc}(S)$\cite{zfis} &$3.630^{(+3)}_{(-17)}$&  & && &&  \\
\tableline
$C_4^{bcc}(S)$\cite{zfis} &3.236(2)&  & && &&  \\
\end{tabular}
\end{table}

\begin{table}
\squeezetable
\caption{Estimates  
of the universal renormalized coupling $g_r$ 
 using the auxiliary function
 $z^{\#}(\beta;S)$ of eq.(\ref{zg})   
 for the spin-$S$ Ising   model series 
 on the sc  and the bcc lattices. They are obtained by  
differential approximants 
 biased with the  modified-ratio  estimates of the 
critical inverse temperatures reported in Table II.
 For comparison, we have also reported other estimates 
 obtained  by simulation methods or from shorter
 series in the recent literature. (For some of them,
 no indication of error is available.) The values marked with 
an asterisk have been obtained  either by renormalization-group
 methods or by high-temperature methods 
 which assume universality and therefore 
they refer to the Ising universality class although, for 
 simplicity, they are reported in the column of the $S=1/2$ results.}
\label{tab6}

\begin{tabular}{lccccccc}
 &  S=1/2 & S=1 & S=3/2 &  S=2&  S=5/2&S=3&  S=$\infty$ \\
\tableline
$g_r^{sc}(S)$ (DA) & 1.40(1) &1.410(6) &1.404(6) &1.414(10) 
&1.415(10) &1.414(10)&1.412(10) \\
\tableline
$g_r^{bcc}(S)$ (DA)& $1.408^{(+1)}_{(-4)}$ &1.409(4) &1.404(3) 
&$1.401^{(+4)}_{(-1)}$ 
&$1.400^{(+5)}_{(-1)}$ &$1.400^{(+5)}_{(-1)}$&$1.398^{(+6)}_{(-1)}$\\ 
\tableline
$ g_r^{bcc}$\cite{mon}& 1.401(8)&&&&&\\
\tableline
$ g_r^{sc}$\cite{mon}& 1.403(7)&&&&&\\
\tableline
$ g^{bcc}_r$\cite{zfis}& 1.459(9)&&&&&\\
\tableline
$ g_r$\cite{campo}& 1.402(2)*&&&&&\\
\tableline
$ g_r$\cite{murnick}& 1.40*&&&&&\\
\tableline
$ g_r$\cite{guida}& 1.411(4)*&&&&&\\
\tableline
\end{tabular}
\end{table}

\begin{table}
\squeezetable

\caption{ Estimates of   
the universal quantity $R^+_{\xi}$ using  
 the auxiliary function $F$ of eq.(\ref{rxipiu})
 for the spin-$S$ Ising   model series 
 on the sc  and the bcc lattices. They are  obtained  by  
 differential approximants
 biased with the critical inverse 
temperatures reported in Table II and 
with the value of $\nu$ obtained in this study.
We have also reported some estimates 
 obtained  by other methods or from shorter
 series in the recent literature. The estimates marked with 
an asterisk are obtained by renormalization-group methods or by 
 high-temperature methods 
that assume universality and therefore 
they refer to the Ising universality class although, for 
 simplicity, they are reported in the column of the $S=1/2$ results.}
\label{tab7}
\begin{tabular}{lccccccc}
 &  S=1/2 & S=1 & S=3/2 &  S=2&  S=5/2&S=3&  S=$\infty$ \\
\tableline
$R^{+sc}_{\xi}(S)$&0.2664(10)&0.2669(12)
&0.2671(11)&0.2673(12)&0.2679(15)&0.2674(10)&0.2673(10) \\
\tableline
$R^{+bcc}_{\xi}(S)$&0.2664(4)&0.2667(3)
&0.2668(3)&0.2669(4)&0.2669(4)&0.2669(4)&0.2670(4) \\
\tableline
$R^{+}_{\xi}$\cite{stauff,liufi2,campo}& 0.2659(4)&&&&&&\\
\tableline
$R^{+}_{\xi}$\cite{haspi}& 0.270(4) &&&&&&\\
\tableline
$R^{+}_{\xi}$\cite{berxi}& 0.270(1)*&&&&&&\\
\tableline
$R^{+}_{\xi}$\cite{bereps}& 0.27* &&&&&&\\
\tableline
$R^{+}_{\xi}$\cite{migr} & 0.284(18)&&&&&&\\
\tableline
$R^{+}_{\xi}$\cite{now} &0.265(6)&&&&&&\\
\end{tabular}
\end{table}

\begin{table}
\squeezetable
\caption{Estimates of the universal ratio $a_{\xi}/a_{\chi}$ 
 for the spin-$S$ Ising   model series 
 on the  bcc lattice.
 For comparison, we have also reported some estimates 
 obtained  by renormalization-group methods. (For some of them,
 no indication of error is available.) 
Although they refer to the Ising universality class, for 
 simplicity, they are reported in the column 
of the $S=1/2$ results.}
\label{tab8}
\begin{tabular}{lccccccc}
 &  S=1/2 & S=1 & S=3/2 &  S=2&  S=5/2&S=3&  S=$\infty$ \\
\tableline
$a^{bcc}_{\chi}(S)$& -0.129(3)&-0.0363(10)&0.0079(8)&0.0307(10)&0.0436(10)&
0.0515(10)& 0.0742(20)\\
\tableline
$a^{bcc}_{\xi}(S)$& -0.100(4)&-0.0279(15)& 0.0061(6)&0.0233(10)& 0.0331(20)&
0.0390(20)& 0.0560(20) \\
\tableline
$a^{bcc}_{\chi}(S)$\cite{georth}& -0.119&-0.034& &0.023& & &  \\
\tableline
$a^{bcc}_{\xi}(S)$\cite{georth}& -0.1085&-0.033& &0.0225& & &  \\
\tableline
$a^{bcc}_{\chi}(S)$\cite{nr90}& -0.13&& && & &  \\
\tableline
$a^{bcc}_{\xi}(S)$\cite{nr90}& -0.11&& && & &  \\
\tableline
\tableline
 $a^{bcc}_{\xi}(S)/a^{bcc}_{\chi}(S)$  
&0.78(5) &0.77(6) &0.77(16) &0.76(6) &0.76(6) &0.76(5)&0.76(5) \\
\tableline
$a_{\xi}/a_{\chi}$ \cite{nr90}&0.85 &&&&&\\
\tableline
$a_{\xi}/a_{\chi}$ \cite{babe}&0.65(5)* &&&&&\\
\tableline
$a_{\xi}/a_{\chi}$ \cite{chre}& 0.65* &&&&&\\
\tableline
\end{tabular}
\end{table}

\newpage

\newpage

\newpage

\begin{figure}[t]
\centerline{\hbox{
\psfig{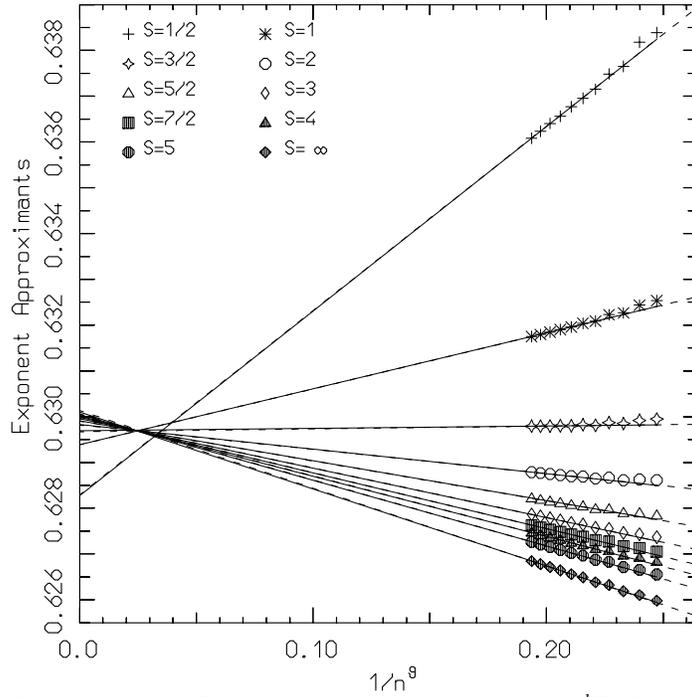}}}
\caption{ Same as in Fig.\ \ref{fig3}, but for the modified-ratio 
 approximant sequences 
$(\nu^{bcc}(S))_n$  of the correlation-length critical exponent, 
as obtained from the expansion of $\xi^{bcc}(\beta;S)$.}
\label{fig5}
\end{figure}

\begin{figure}[b]
\centerline{\hbox{
\psfig{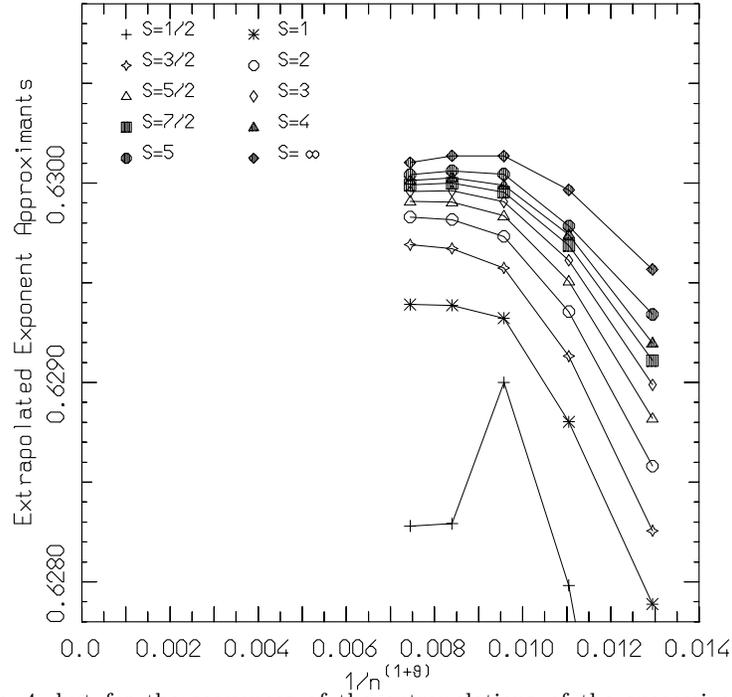}}}
\caption{ Same as in Fig.\ \ref{fig4}, but for 
the sequences of the extrapolations of the successive odd pairs of
modified-ratio  approximants   $(\nu^{bcc}(S))_n$ vs.
 $1/n^{1+\theta}$.}
\label{fig6} 
\end{figure}

\newpage

\begin{figure}[t]
\centerline{\hbox{
\psfig{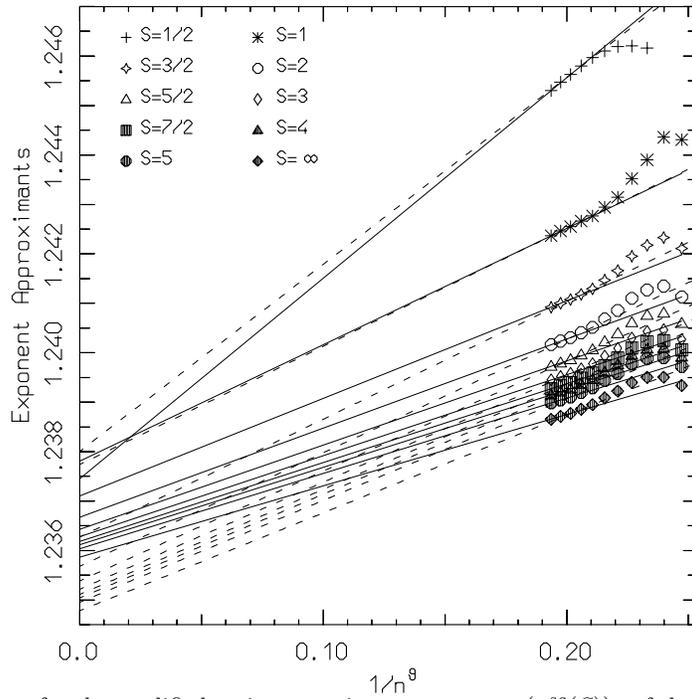}}}
\caption{ Same as in Fig.\ \ref{fig3}, but for the modified-ratio 
 approximant sequences
 $(\gamma^{sc}(S))_n$ of the susceptibility critical exponent 
as obtained from $\chi^{sc}(\beta;S)$ using  eq.(\ref{gammazinn}).}
\label{fig7}
\end{figure}

\begin{figure}[b]
\centerline{\hbox{
\psfig{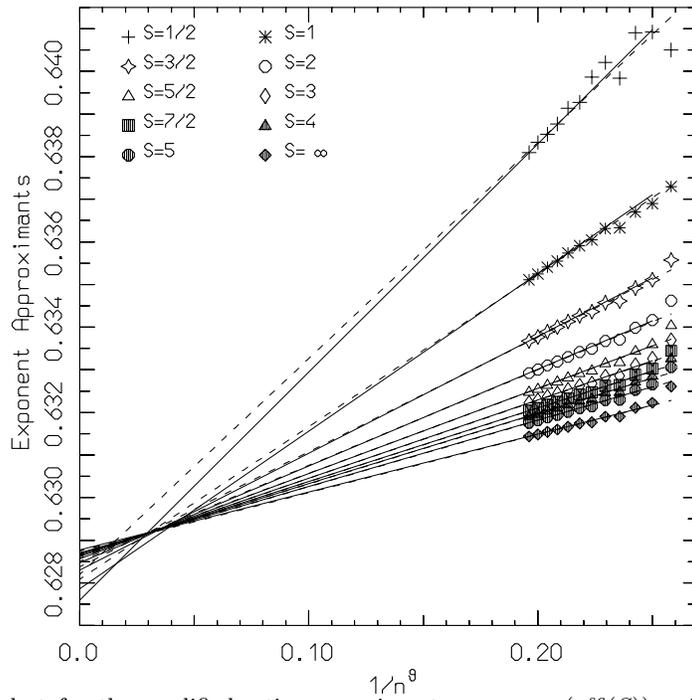}}}
\caption{Same as in Fig.\ \ref{fig3}, but for the  modified-ratio  
approximant  sequences 
$(\nu^{sc}(S))_n$  of the correlation-length critical exponent, 
as obtained from $\xi^{sc}(\beta;S)$.}
\label{fig8}
\end{figure}

\newpage
\begin{figure}[t]
\centerline{\hbox{
\psfig{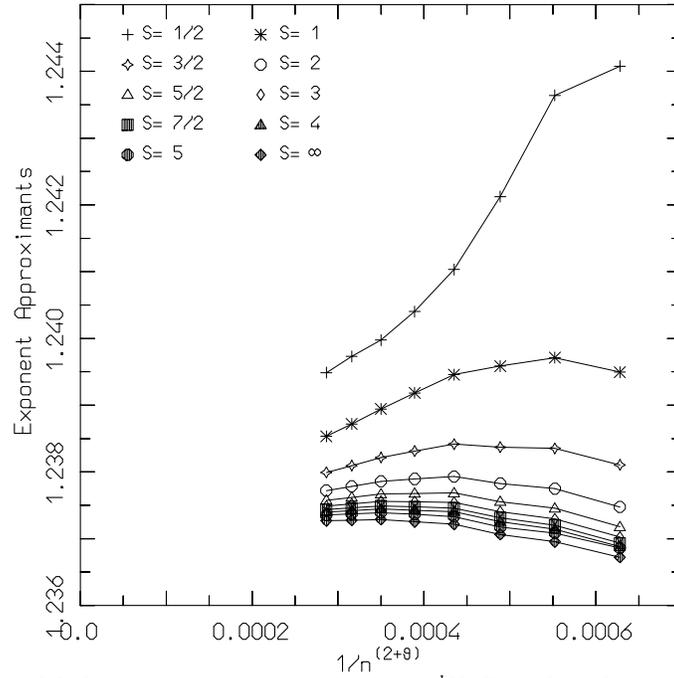}}}
\caption{The {\it directly} biased modified-ratio   
approximant sequences  $(\hat \gamma^{bcc}(S))_n$
 plotted vs. $1/n^{2+\theta}$. 
 They are obtained from 
eq.(\ref{gammazinnth}),  using as a bias 
the value of $\theta$
 in order to reduce the influence of the confluent corrections to scaling. 
 The continuous lines are only
 guides to the eye.}
\label{fig9}
\end{figure}
 
\begin{figure}[b]
\centerline{\hbox{
\psfig{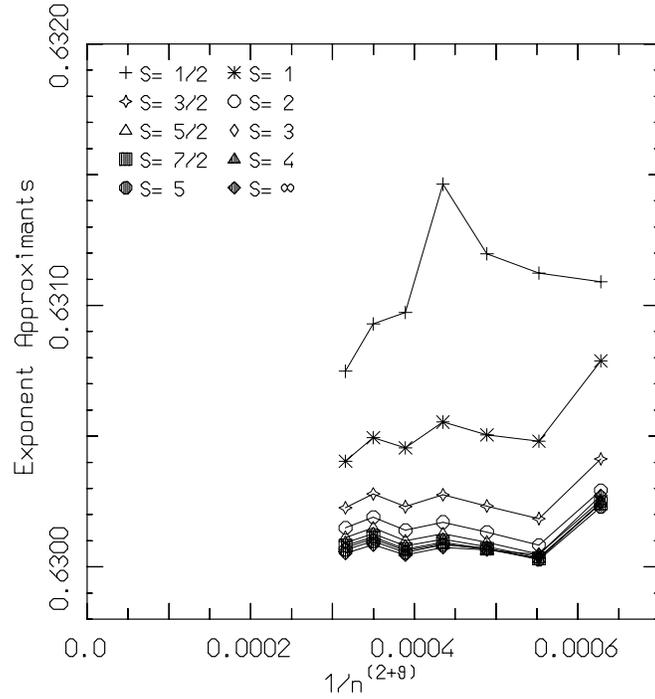}}}
\caption{ Same as Fig.\ \ref{fig8}, but for the {\it directly} biased 
modified-ratio  approximant 
sequences  $(\hat \nu^{bcc}(S))_n$ plotted vs. $1/n^{2+\theta}$.}
\label{fig10}
\end{figure}

\newpage
\begin{figure}[t]
\centerline{\hbox{
\psfig{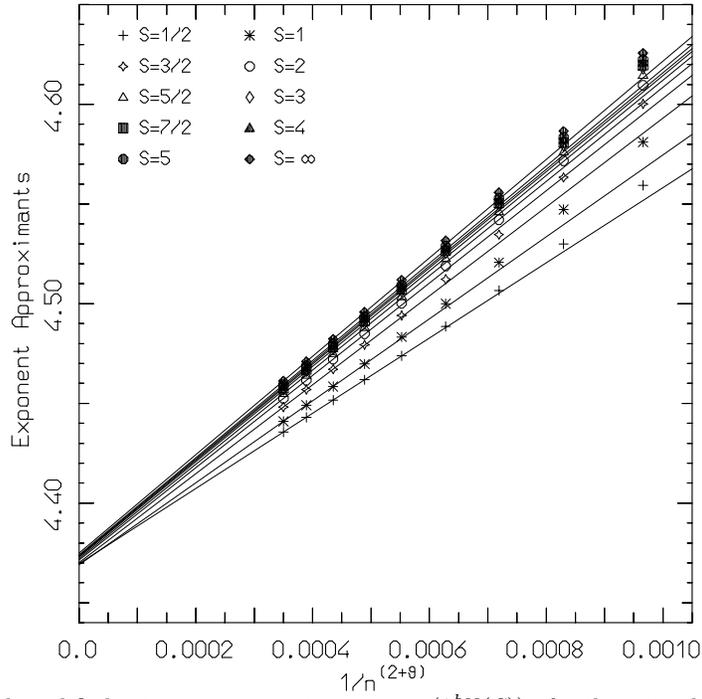}}}
\caption{ The {\it directly} biased modified-ratio  approximant sequences  
$(\hat \gamma_4^{bcc}(S))_n$ for the critical exponent 
 of  $\chi_4^{bcc}(\beta;S)$ plotted vs. $1/n^{2+\theta}$.
In order to keep the figure readable
 we have indicated only the extrapolations 
of the odd approximant  subsequences. }
\label{fig11}
\end{figure}

\begin{figure}[b]
\centerline{\hbox{
\psfig{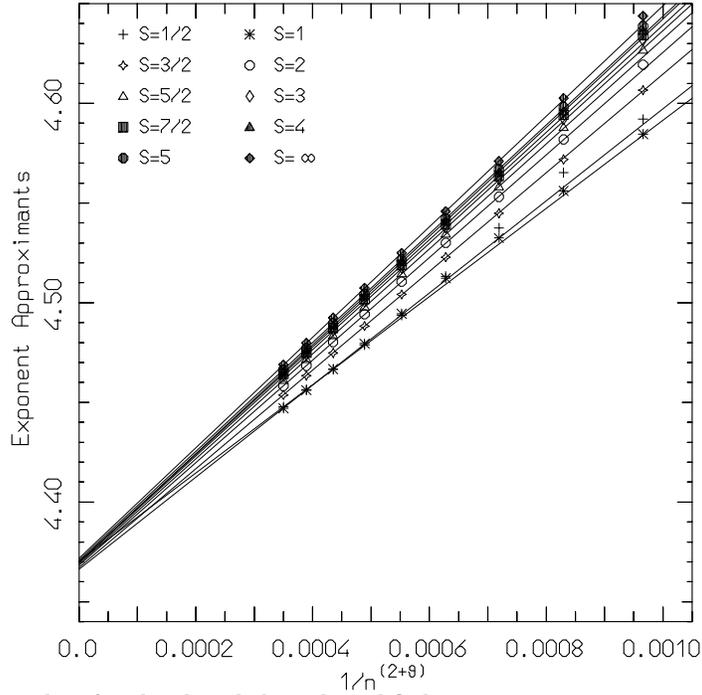}}}
\caption{ Same as in Fig.\ \ref{fig11}, but for 
 the  {\it directly} biased modified-ratio  approximant sequences  
$(\hat \gamma_4^{sc}(S))_n$ for the critical exponent 
 of  $\chi_4^{sc}(\beta;S)$ plotted vs. $1/n^{2+\theta}$.}
\label{fig12}
\end{figure}

\newpage
\begin{figure}[t]
\centerline{\hbox{
\psfig{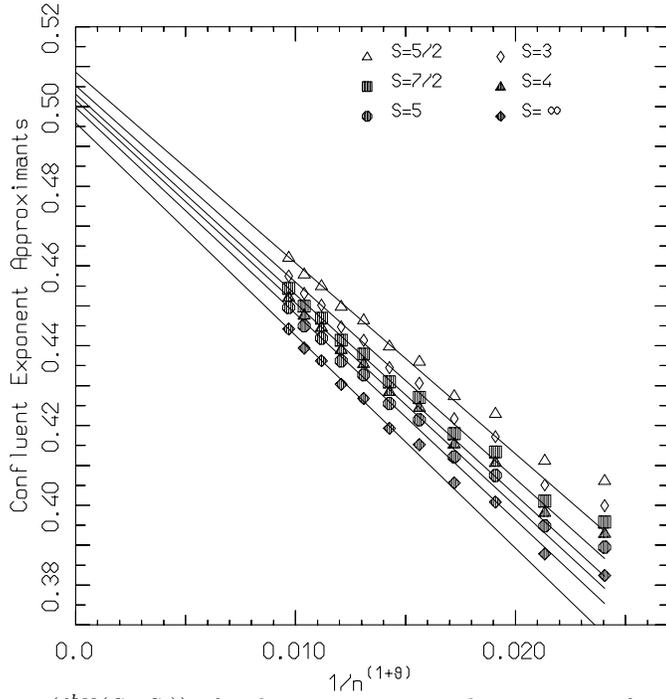}}}
\caption{Approximant sequences  
$(\theta^{bcc}(S_1,S_2))_n$ for the  correction-to-scaling exponent 
  $\theta$ as obtained using eq.(\ref{thseq}) for fixed $S_1=1/2$ 
and $S_2=5/2,3,7/2,4,5,\infty$. The symbols refer to the values
 of  $S_2$.
The sequences are plotted vs. $1/n^{1+\theta}$, with $\theta= 0.504$. 
In order to keep the figure readable
 we have indicated only the extrapolations 
of the odd approximant  subsequences.}
\label{fig13}
\end{figure}



\begin{figure}[b]
\centerline{\hbox{
\psfig{figure=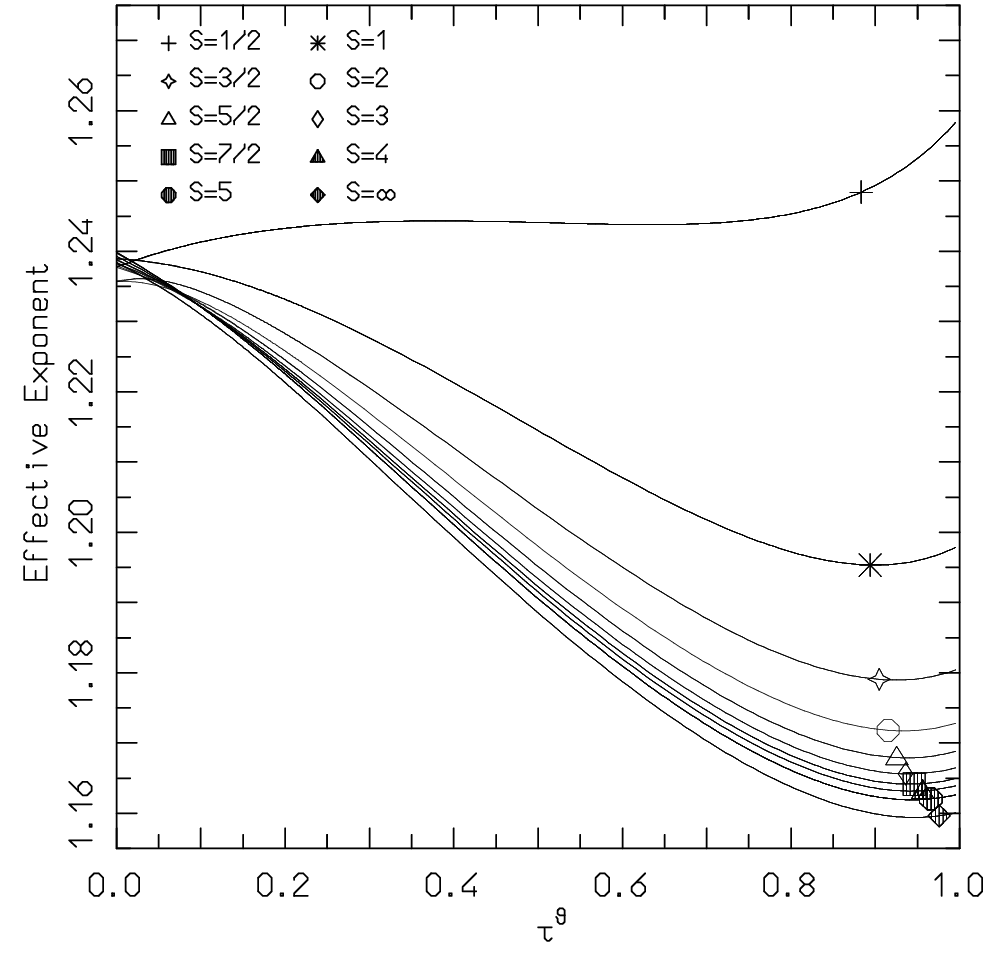,height=3.60in}}}
\caption{Highest order simplified-differential approximants
 of the effective  exponent $\gamma^{bcc}_{eff}(\beta;S)$
of the susceptibility $\chi^{bcc}(\beta;S)$ as defined by eq.(\ref{effg}).
 For each value of the spin $S$ the effective exponent is plotted vs.  
$\tau^{bcc}(S)^{\theta}=(1-\beta/\beta^{bcc}_c(S))^{\theta}$.
As indicated by the symbols attached to them, 
the curves refer, from the highest downwards, to  
the spin values $S=1/2,1,3/2,2,5/2,3,7/2,4,5,\infty$. }
\label{fig14}
\end{figure} 


\newpage

\begin{figure}[t]
\centerline{\hbox{
\psfig{figure=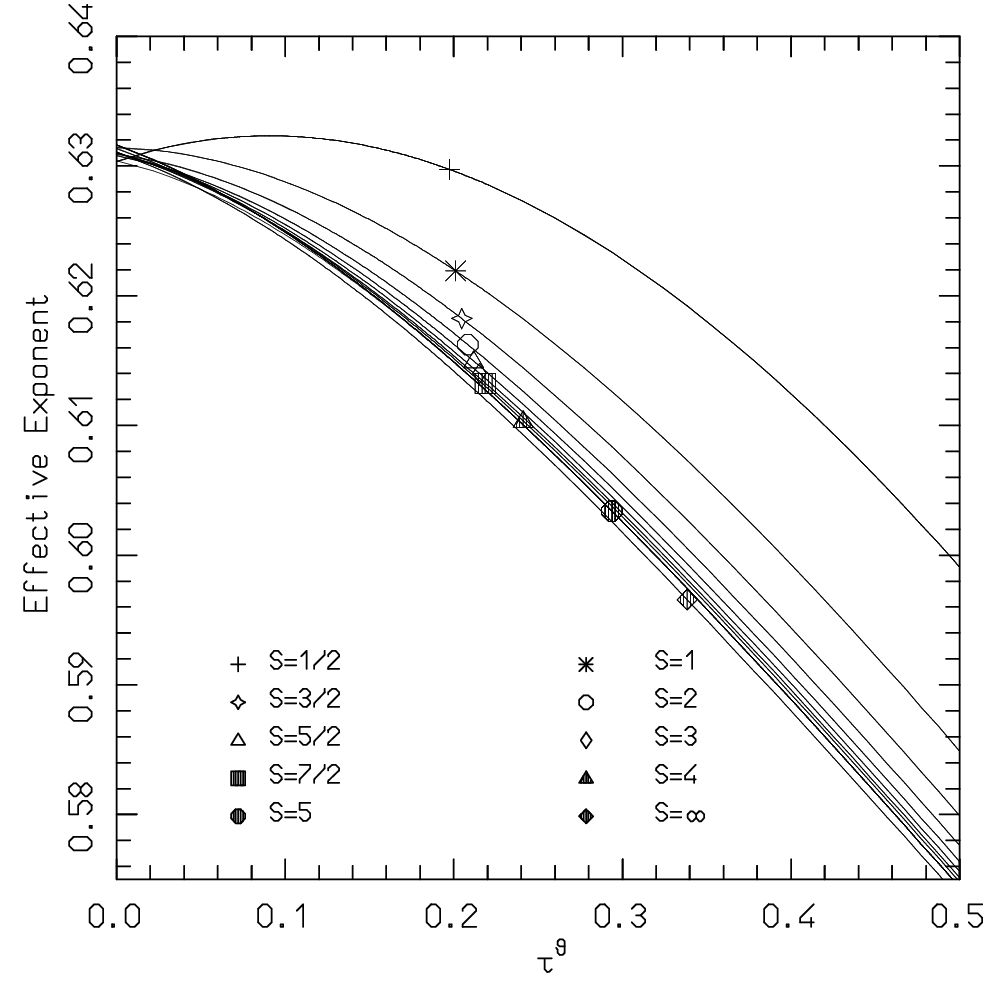,height=3.60in}}}
\caption{Highest order simplified-differential approximants
 of the effective  exponent $\nu^{bcc}_{eff}(\beta;S)$
of the correlation length $\xi^{bcc}(\beta;S)$ as defined by eq.(\ref{effg}).
 For each value of the spin $S$ the effective exponent is plotted vs.  
$\tau^{bcc}(S)^{\theta}=(1-\beta/\beta^{bcc}_c(S))^{\theta}$.
As indicated by the symbols attached to them, 
the curves refer, from the highest downwards, to  
the spin values $S=1/2,1,3/2,2,5/2,3,7/2,4,5,\infty$. }
\label{fig15}
\end{figure} 


\begin{figure}[b]
\centerline{\hbox{
\psfig{figure=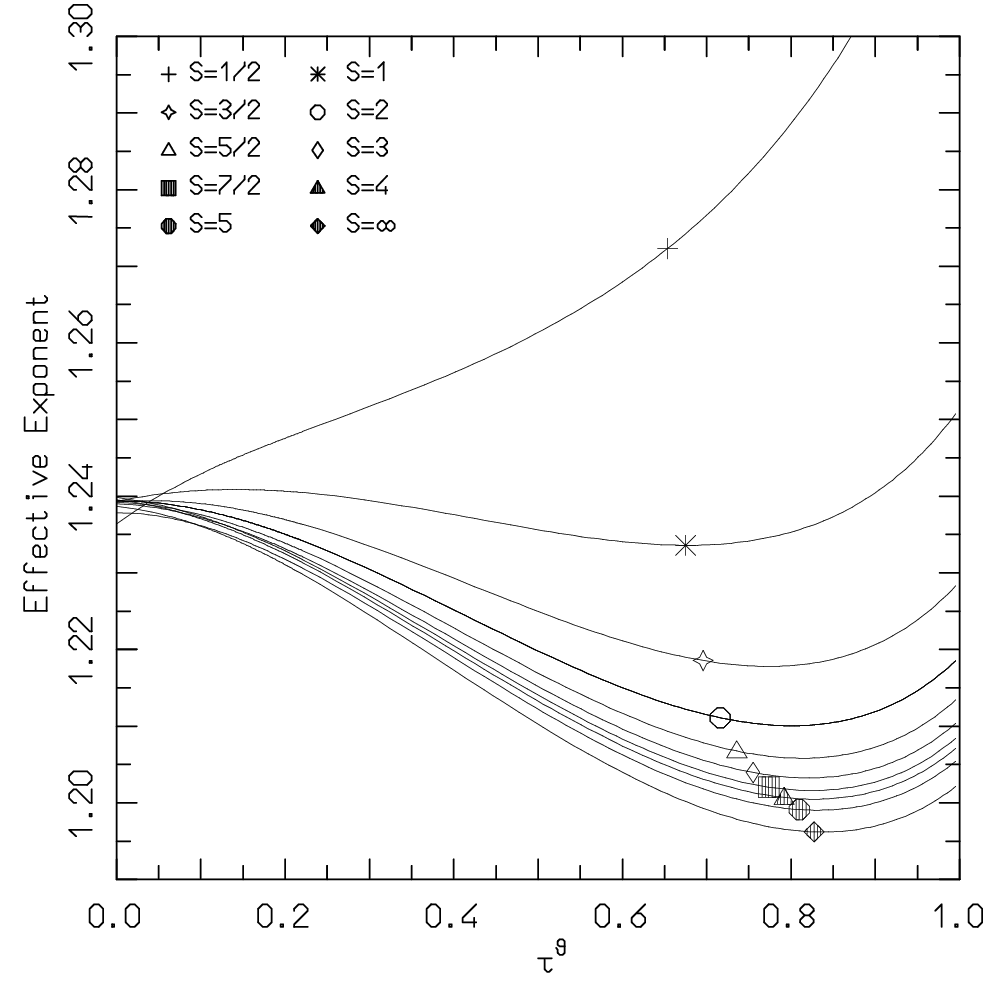,height=3.60in}}}
\caption{Highest order simplified-differential  approximants of the effective  
exponent $\gamma^{sc}_{eff}(\beta;S)$
of the susceptibilty $\chi^{sc}(\beta;S)$ as defined by eq.(\ref{effg}).
 For each value of the spin $S$ the effective exponent is plotted vs.  
$\tau^{sc}(S)^{\theta}=(1-\beta/\beta^{sc}_c(S))^{\theta}$.
As indicated by the symbols attached to them, 
the curves refer, from the highest downwards, to  
the spin values $S=1/2,1,3/2,2,5/2,3,7/2,4,5,\infty$. }
\label{fig16}
\end{figure} 

\newpage

\begin{figure}[t]
\centerline{\hbox{
\psfig{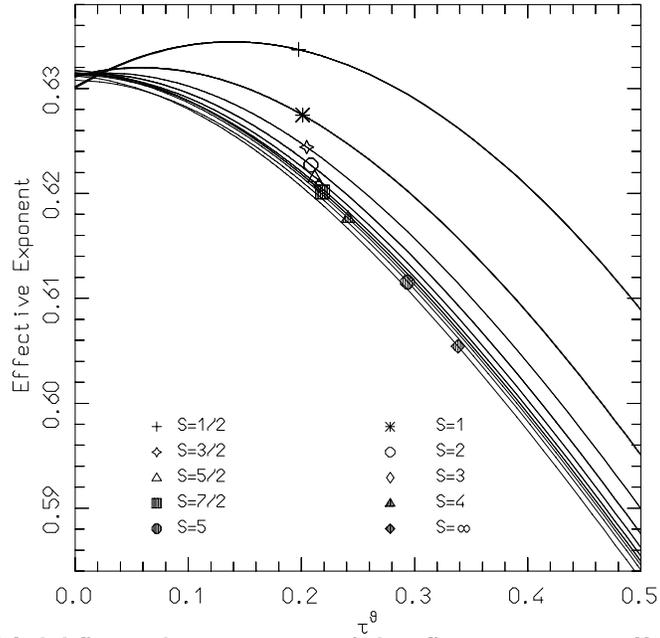}}}
\caption{Highest order simplified-differential approximants of the effective  
exponent $\nu^{sc}_{eff}(\beta;S)$
of the correlation length $\xi^{sc}(\beta;S)$ as defined by eq.(\ref{effg}).
 For each value of the spin $S$ the effective exponent is plotted vs.  
$\tau^{sc}(S)^{\theta}=(1-\beta/\beta^{sc}_c(S))^{\theta}$.
As indicated by the symbols attached to them, 
the curves refer, from the highest downwards, to  
the spin values $S=1/2,1,3/2,2,5/2,3,7/2,4,5,\infty$. }
\label{fig17}
\end{figure}


\begin{figure}[b]
\centerline{\hbox{
\psfig{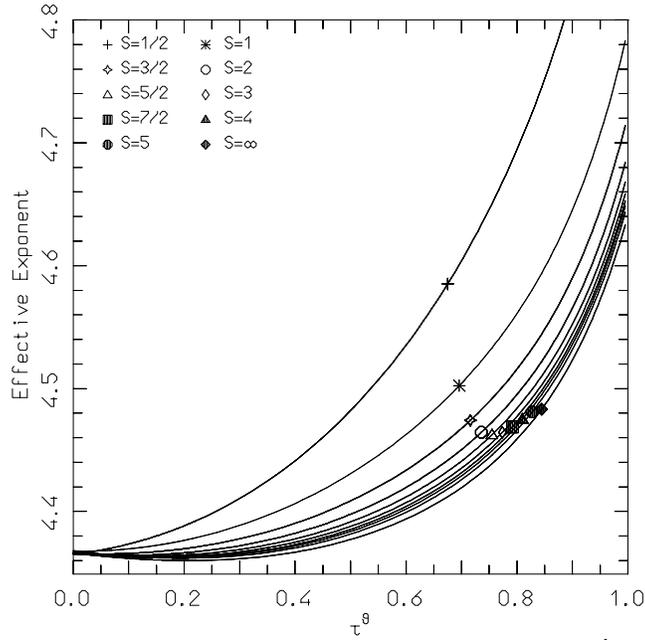}}}
\caption{Highest order simplified-differential approximants of 
 the effective exponent  $\gamma^{bcc}_{4eff}(\beta;S)$
 computed from $\chi^{bcc}_4(\beta;S)$.  
 For each value of the spin $S$ the effective exponent is plotted versus 
 the corresponding reduced inverse temperature 
$\tau^{bcc}(S)^{\theta}=(1-\beta/\beta^{bcc}_c(S))^{\theta}$ }
\label{fig18}
\end{figure}

\newpage


\begin{figure}[t]
\centerline{\hbox{
\psfig{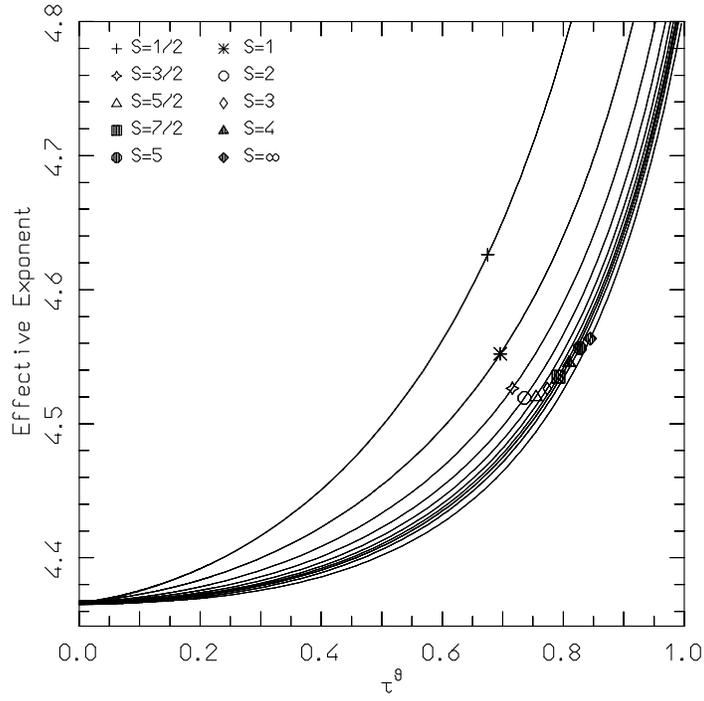}}}
\caption{ Same as Fig.\ \ref{fig18}, but for   
the effective exponent $\gamma^{sc}_{4eff}(\beta;S)$
 computed from $\chi_4^{sc}(\beta;S)$. }
\label{fig19}
\end{figure}

\begin{figure}[b]
\centerline{\hbox{
\psfig{figure=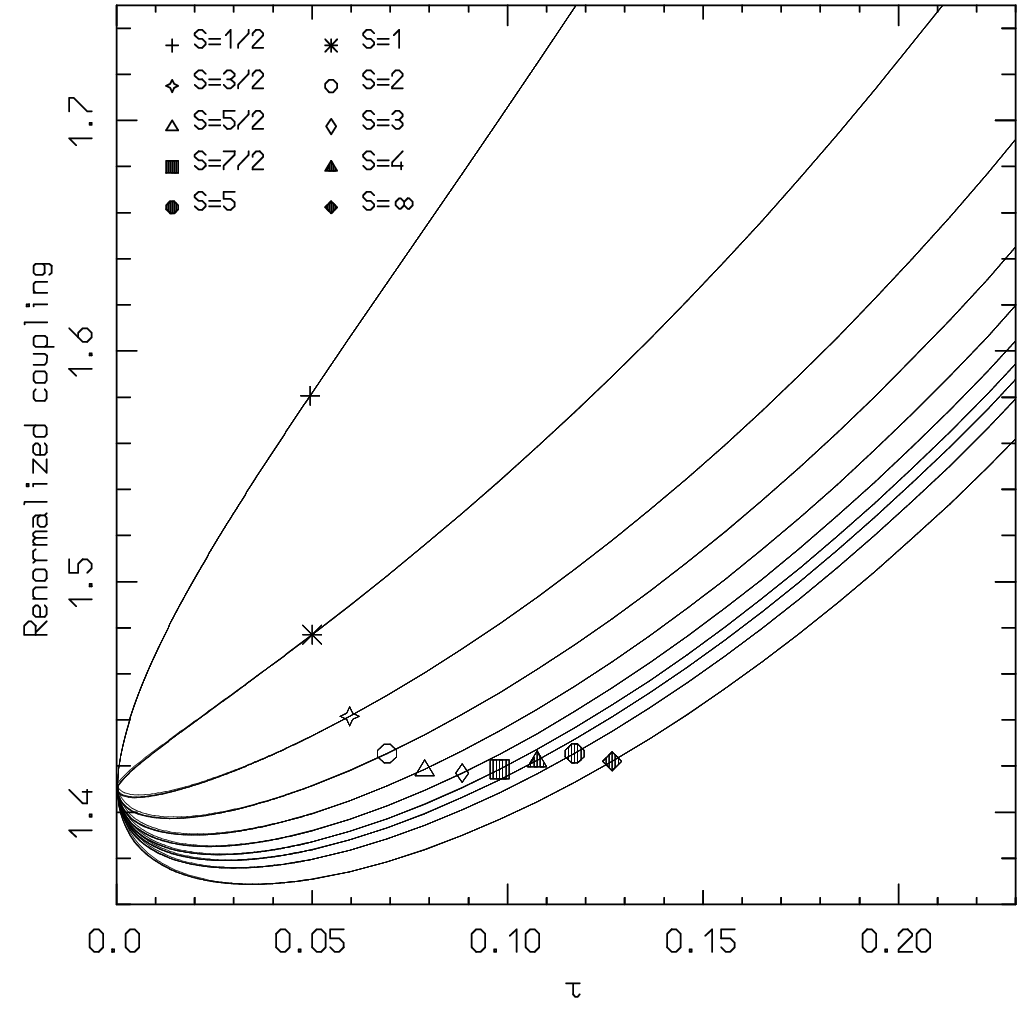,height=3.60in}}}
\caption{Highest order simplified differential approximants of the 
effective dimensionless renormalized coupling constant $g_r(\beta;S)$, 
as obtained 
from the bcc lattice series.
 The effective coupling is computed from the auxiliary function  
$z^{bcc}(\beta;S)$ defined in eq.(\ref{zg}) and is plotted 
vs. $\tau^{bcc}(S)=1-\beta/\beta^{bcc}_c(S)$.   }
\label{fig20}
\end{figure} 

\newpage

\begin{figure}[t]
\centerline{\hbox{
\psfig{figure=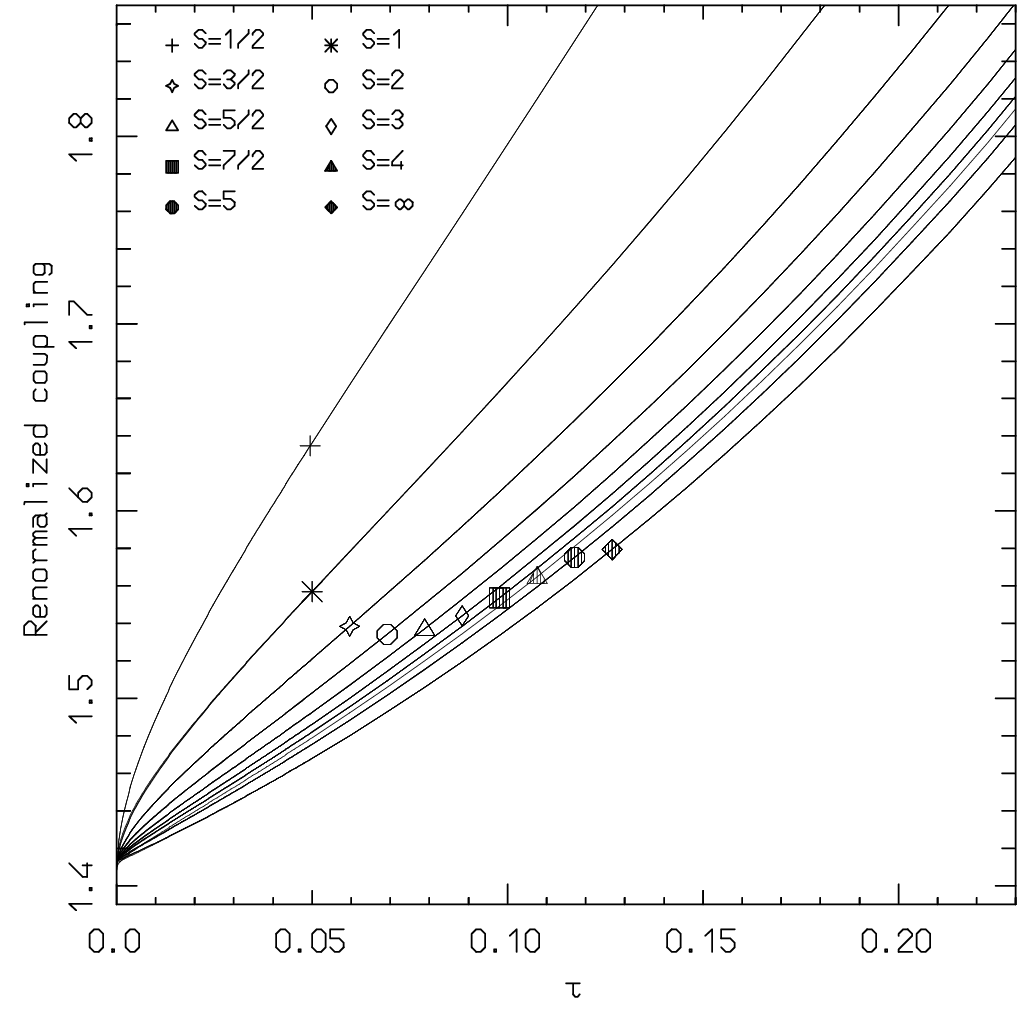,height=3.60in}}}
\caption{ Highest order simplified differential 
approximants of the effective  
dimensionless renormalized coupling constant $g_r(\beta;S)$ as obtained 
from the sc lattice series.
 The effective coupling is computed from the auxiliary function  
$z^{sc}(\beta;S)$ defined by eq.(\ref{zg}) and is plotted 
vs. $\tau^{sc}(S)=1-\beta/\beta^{sc}_c(S)$.   }
\label{fig21}
\end{figure} 


\begin{references}
\bibitem{fam} The fact that for a simple antiferromagnet in zero field 
the susceptibility has a critical energy-like behavior was first pointed
 out by Sykes and Fisher\cite{syfi} and observed experimentally in
 Ref.\cite{wowy}.
\bibitem{afc} More complex behaviors can appear in presence of 
antiferromagnetic couplings, suggesting\cite{gri}
  possible ways for defining a non-trivial continuum limit.
\end{references}
\end{document}